\documentclass[%
 reprint,
 amsmath,amssymb,
 aps,
]{revtex4-2}

\usepackage{graphicx}
\usepackage{amssymb}
\usepackage{amsmath}
\usepackage{epstopdf}
\usepackage{graphics}
\usepackage[caption=false,subrefformat=parens,labelformat=parens]{subfig}
\usepackage{float}
\usepackage{color}
\usepackage[colorlinks=true]{hyperref}
\usepackage{wasysym}
\usepackage{xcolor}
\usepackage{bm}

\allowdisplaybreaks[1]

\newcommand{\bra}[1]{\langle #1|}					
\newcommand{\ket}[1]{|#1\rangle}					
\newcommand{\braket}[2]{\langle #1 | #2 \rangle}	

\newcommand{\abs}[1]{\left| #1 \right|} 
\newcommand{\appropto}{\mathrel{\vcenter{
  \offinterlineskip\halign{\hfil$##$\cr
    \propto\cr\noalign{\kern2pt}\sim\cr\noalign{\kern-2pt}}}}}

\makeatletter
\renewcommand\frontmatter@abstractwidth{\dimexpr\textwidth-2in\relax}
\makeatother

\begin{document}
\title{Entanglement distribution with minimal memory requirements using time-bin photonic qudits}
\author{Yunzhe Zheng}
\affiliation{QuTech and Kavli Institute of Nanoscience, Delft University of Technology, 2628 CJ, Delft, The Netherlands}
\author{Hemant Sharma}
\affiliation{QuTech and Kavli Institute of Nanoscience, Delft University of Technology, 2628 CJ, Delft, The Netherlands}
\author{Johannes Borregaard}
\affiliation{QuTech and Kavli Institute of Nanoscience, Delft University of Technology, 2628 CJ, Delft, The Netherlands}
\date{\today}
\begin{abstract}
Generating multiple entangled qubit pairs between distributed nodes is a prerequisite for a future quantum internet. To achieve a practicable generation rate, standard protocols based on photonic qubits require multiple long-term quantum memories, which remains a significant experimental challenge. In this paper, we propose a novel protocol based on $2^m$-dimensional time-bin photonic qudits that allow for the simultaneous generation of multiple ($m$) entangled pairs between two distributed qubit registers and outline a specific implementation of the protocol based on cavity-mediated spin-photon interactions. By adopting the qudit protocol, the required qubit memory time is independent of the transmission loss between the nodes in contrast to standard qubit approaches. As such, our protocol can significantly boost the performance of near-term quantum networks.
\end{abstract}
\maketitle


\section{Introduction}
Quantum networks bring new opportunities for secure communication~\cite{Lo2014,Pirandola2020}, distributed sensing~\cite{Komar2014,Khabiboulline2019,Guo2020}, and distributed  quantum computing~\cite{Meter2016,wehner2018}. The ability to faithfully transmit quantum information over long distances is a key prerequisite for the construction of large-scale quantum networks. Quantum teleportation~\cite{Pirandola2015} provides a means to overcome transmission loss and noise given that high-quality entanglement can be created between the sender and the receiver. As such, there have been significant efforts for creating entanglement between remote qubit systems across a plethora of physical systems including trapped ions~\cite{Simon2003}, diamond defect centers~\cite{Hensen2015,Pompili2021}, neutral atoms~\cite{Langenfeld2021,Lago-Rivera2021} and quantum dots~\cite{Delteil2016}. 

In canonical entanglement generation protocols, spin-photon entanglement is first created and then extended to spin-spin entanglement by either a photonic Bell measurement~\cite{Simon2003,Hensen2015,Pompili2021,Lago-Rivera2021,Delteil2016} or a direct spin-photon interaction~\cite{Langenfeld2021}. Successful entanglement generation is heralded by the detection of transmitted photons, and the quantum state of the spin qubits need to stay coherent for at least the time of one entanglement generation attempt, which is generally set by the signaling time between the nodes. 

Key functionalities in a quantum network such as entanglement purification~\cite{Bennett1996,Deutsch1996} and multi-qubit state teleportation~\cite{Fowler2010,Munro2015,Muralidharan2016} require the availability of more than a single entangled qubit pair to be executed. The latter, in particular, when quantum information encoded across multiple physical qubits in a quantum error correcting code has to be transmitted since all physical qubits of the code has to be transmitted simultaneously in order to perform error correction locally. 

Multi-qubit memories are therefore necessary, and the general approach so far has been to consider the generation of multiple entangled pairs either sequentially~\cite{Childress2005,Kalb2017} or in parallel~\cite{Liang2009,Munro2015} using the same photonic qubit based protocols as for the generation of a single entangled pair. High transmission loss, however, leads to much more demanding requirements for the coherence time of the quantum memories than for the single pair setup. For a feasible generation rate, successfully entangled pairs have to be stored in quantum memories while they wait for remaining pairs to entangle successfully. As a result, the required memory time increases as the inverse of the photon transmission probability, which decreases exponentially with the distance between the nodes for standard optical fiber propagation. To add insult to injury, the continued entanglement attempts of neighboring qubits may further decrease the memory time of a successful pair due to unwanted crosstalk~\cite{Kalb2018}. 

\begin{figure} [t]
    \centering
    \includegraphics[width = 8.6cm]{ 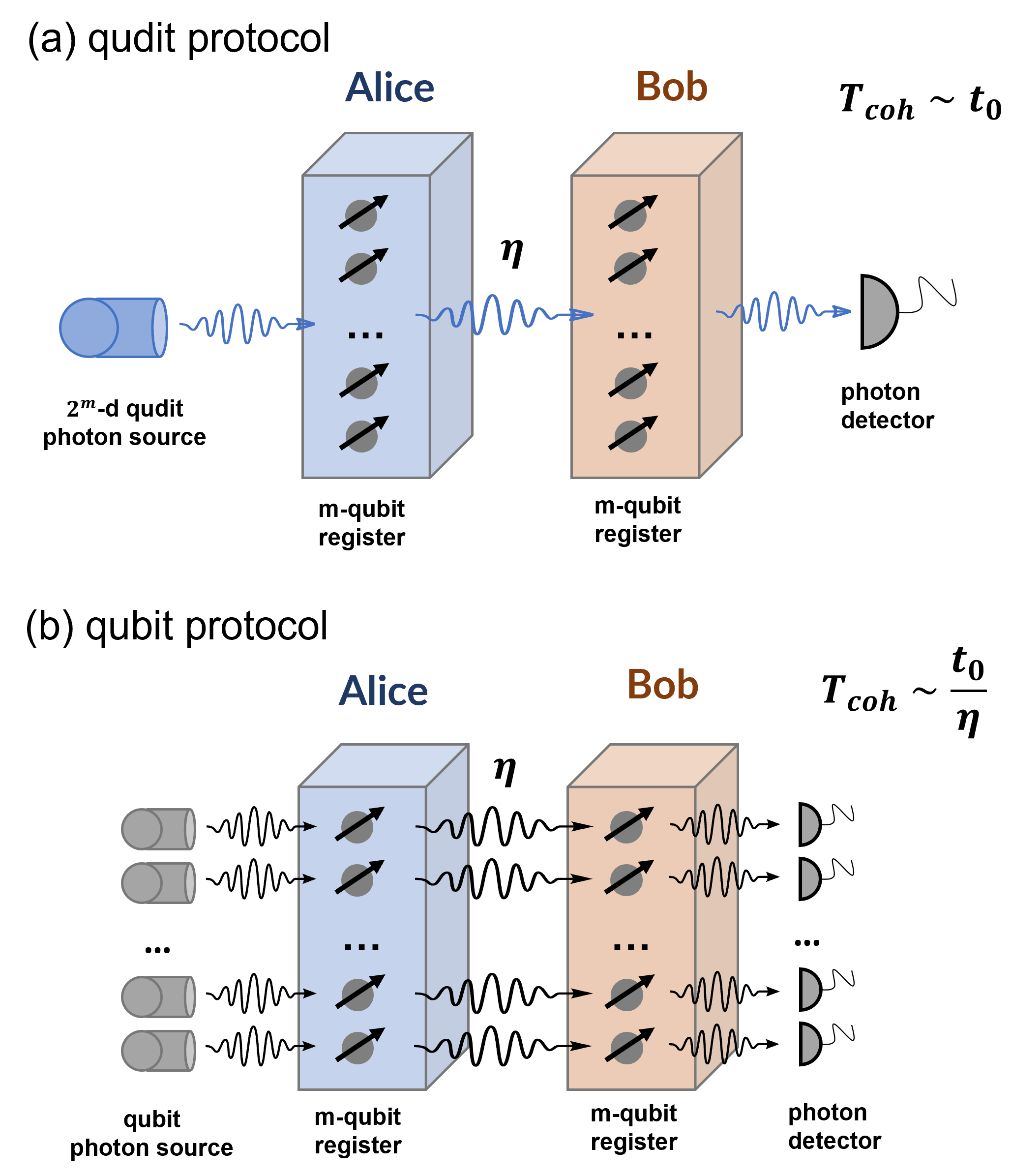}
    \caption{(a) The qudit protocol for simultaneous generation of multiple entangled pairs. First, entanglement between a single photonic qudit and all memory-qubits on Alice's side is generated. A $2^m$-dimensional photonic qudit is required for $m$ memory-qubits. The qudit is transmitted to Bob's register and interacts with all memory-qubits thereby extending the entanglement to this register. A final measurement decouples the photonic state from the memory-qubits and heralds the simultaneous generation of all entangled pairs. The required coherence time of the memory qubits is $T_{coh}\sim t_0$ where $t_{0}$ is the time of one attempt. (b) A comparative qubit-based protocol for the generation of multiple entangled pairs between two qubit registers. Entanglement is first created between multiple photonic qubits and the memory-qubits of Alice's register in a parallel fashion. The photons (each entangled with one memory qubit) are transmitted to Bob's register to interact with his memory-qubits and measurements decouple the photonic states from the memory-qubit states and herald the successful generation of Bell pairs. To efficiently generate multiple pairs, successfully generated pairs have to be stored while the remaining unsuccessful pairs are re-attempted. This leads to a required coherence time of the spin-gates of $T_{coh}\sim t_0/\eta$ where $\eta$ is the transmission probability.}
    \label{fig:1}
\end{figure}

In this article, we propose a fundamentally different scheme for the generation of multiple entangled pairs between two distant qubit registers, which exploits the use of high-dimensional photonic qudit states. Specifically, we show that a single photonic time-bin qudit in dimension $d=2^m$ allows for the heralded and simultaneous generation of $m$ entangled pairs between two distant multi-qubit registers. Photonic qudit encodings have previously been considered for quantum networks~\cite{Bacco2021} and quantum repeaters~\cite{Muralidharan2018} due to their higher information capacity ~\cite{Imany2019}, which can lead to higher loss tolerance and more efficient quantum key generation~\cite{Bell2022}.  In our qudit-based protocol, the required coherence time of the qubit memories is independent of the transmission probability and amounts only to the time of a single entanglement generation attempt. Furthermore, the rate of entanglement generation in our protocol is more robust to transmission loss between the distant registers than other qubit approaches. Our scheme opens up new opportunities for the generation of multiple high-fidelity entangled qubit pairs in an extended quantum network, where transmission loss between the nodes becomes a limiting factor both on generation rate and quantum memory requirements. 
 \section{High-level Protocol}

A high-level sketch of our protocol is shown in Fig.~\ref{fig:1}(a) to be compared with a similar qubit-based protocol in Fig.~\ref{fig:1}(b). We note that multiple other qubit-based protocols exist~\cite{Simon2003,Hensen2015,Pompili2021,Lago-Rivera2021,Delteil2016} but they all share the same fundamental scaling of success probability and necessary coherence time with the transmission probability. Without loss of generality, we can therefore compare our qudit protocol with the specific qubit-based protocol considered here. 

We will first describe our qudit protocol at an abstract level to convey the general idea and then proceed with discussion of specific implementations in sec.~\ref{secIII}. In the first step, a photonic qudit is generated. We specifically consider a photonic time-bin qudit prepared in an equal superposition across $2^m$ modes as described by the state
\begin{equation} \label{eq:equation1}
\ket{\psi}_{ph}=\frac{1}{2^{m/2}}\sum_{l=0}^{2^m-1}\ket{l}_{ph},
\end{equation}
where $\ket{l}_{ph}$ denotes the state where the photon is located in the $l$'th time-bin.

\begin{figure}[t]
    \centering
    \includegraphics[width = 8.6cm]{ 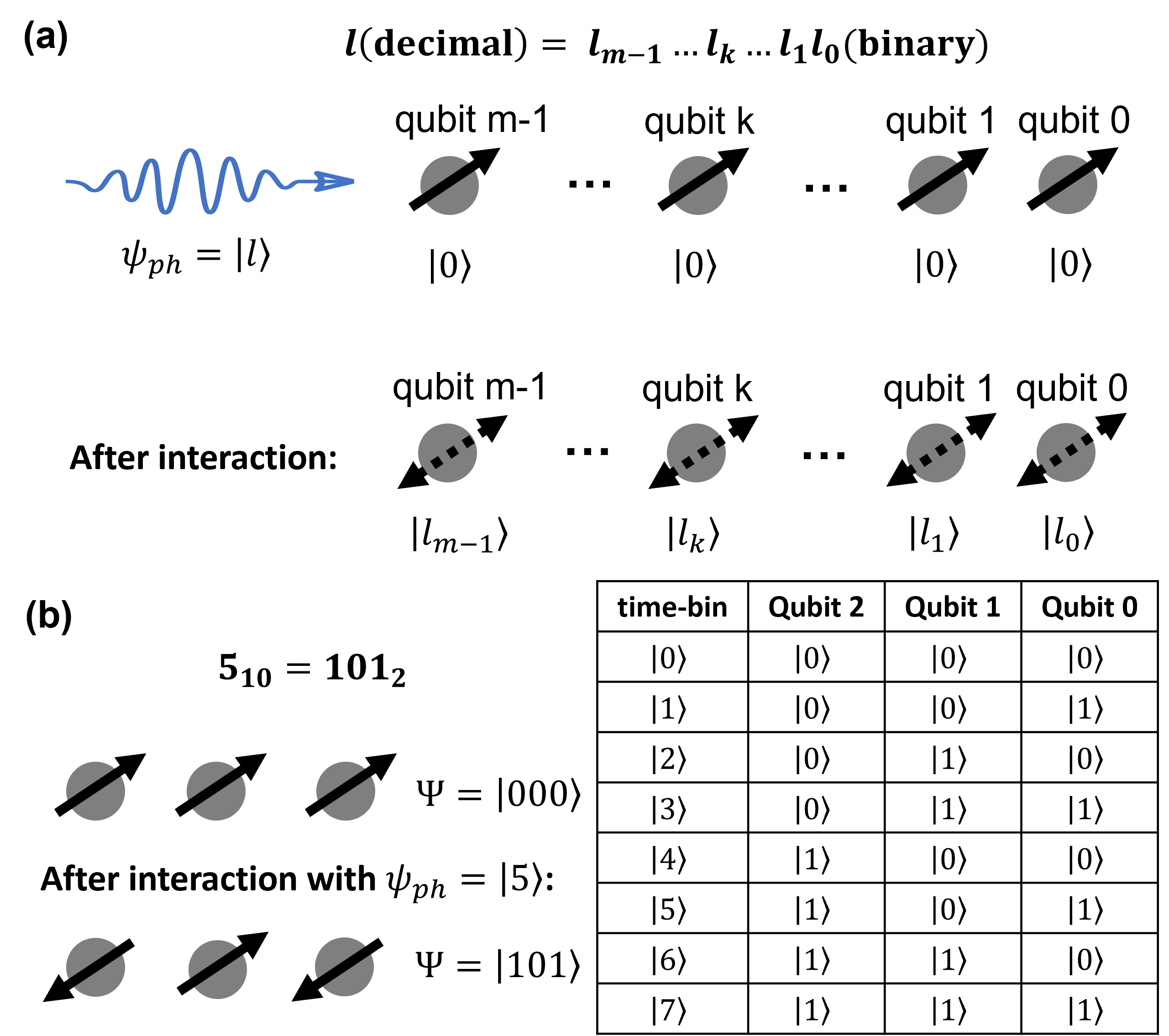}
    \caption{(a) Interaction between the photonic qudit and an $m$-qubit register. The $m$-qubit register is initialized at $\ket{0}^{\otimes m}$. If the photonic qudit is in state $\ket{l}$ ($l = 0, 1,...,2^m-1$) and the decimal value of $l$ can be represented in binary $[l]_{10}=[l_{m-1}...l_{k}...l_1 l_0]_{2}$, the photonic qudit will flip the n-qudit register to state $\ket{l_{m-1}...l_{k}...l_1 l_0}$. (b) Example for a photonic state $\ket{5}$ and a three-qubit register with a value table given for the $m=3$ case.}
    \label{fig:2}
\end{figure}

The photon state $\ket{\psi}_{ph}$ will then interact with the first $m$-qubit register (Alice) which is initialized in state $\ket{0}^{\otimes m}$ in the following way: If the photon is in state $\ket{l}$ where the binary representation of the decimal value $l$ is $[l]_{10}=[l_{m-1}l_{m-2}...l_{0}]_2$ ($l_{i}\in\{0,1\}$), the photon will interact with the $i$'th qubit of the register and flip it from $\ket{0}$ to $\ket{1}$ if $l_i=1$. If $l_i=0$, the qubit remains in state $\ket{0}$. In section~\ref{secIII}, we will outline how such an interaction can be achived with optical switches and cavity-based scattering gates~\cite{Duan2004, Bhaskar2020}. This interaction scheme results in the following transformation
\begin{equation}
    \ket{l}_{ph}\otimes \ket{0}_A^{\otimes m} \xrightarrow[]{} \ket{l}_{ph}\otimes \ket{1_l}_A,
\end{equation}
where $\ket{1_l}_A = \ket{l_{m-1}}\ket{l_{m-2}}...\ket{l_1} \ket{l_0}_A$ as shown in Fig.~\ref{fig:2}(a). For example, a state $\ket{5}_{ph}\otimes\ket{000}$ would transform to
\begin{equation}
    \ket{5}_{ph}\otimes \ket{000} \xrightarrow[]{} \ket{5}_{ph}\otimes \ket{101}
\end{equation}
as illustrated in Fig.~\ref{fig:2}(b). For the input state in Eq.~(\ref{eq:equation1}), the collective state after the photon interacting with Alice's register is 
\begin{equation} \label{eq:spinqudit1}
\ket{\Psi_1}=\frac{1}{2^{m/2}}\sum_{l=0}^{2^m-1}\ket{l}_{ph}\ket{1_l}_A,
\end{equation}
which is a maximally entangled state between the $2^m$ dimensional time-bin photon and the $m$-qubit register.

In the next step, the photonic qudit is sent to Bob's register by means of direct transmission. In practice, the photon most likely will get lost during the transmission, but we will continue our description assuming that the photon is successfully transmitted since we will herald the protocol on a final detection of the photon on Bob's side. 

On Bob's side, the photonic qudit will interact with his $m$-qubit register in the same way as described for Alice's register. Thus, the state of both qubit registers (Alice and Bob) and the photonic qudit after the interaction will be
\begin{equation} \label{eq:spinqudit2}
\ket{\Psi_2}=\frac{1}{2^{m/2}}\sum_{l=0}^{2^m-1}\ket{l}_{ph}\ket{1_l}_A\ket{1_l}_B,
\end{equation}
where $\ket{1_l}_B = \ket{l_{m-1}}\ket{l_{m-2}}...\ket{l_1} \ket{l_0}_B$ denotes the state of Bob's register. 

We now wish to perform a measurement that confirms that the photonic qudit was transmitted and heralds the simultaneous generation of entanglement between the two qubit registers. The measurement should confirm the arrival of the photon without extracting its time-bin information in order to avoid collapsing the state of the two $m$-qubit registers to a product state. To this end, Bob can perform a generalized X-basis measurement on the photonic qudit, which amounts to projecting the photon state on to the time-bin Fourier basis states defined as
\begin{equation}
\ket{\phi_l}_{ph}=\frac{1}{\sqrt{2^m}}\sum_{k=0}^{2^m-1}e^{2i\pi kl/2^m}\ket{k}_{ph},    
\end{equation}
where $l = 0, 1,...,2^m-1$. Such a measurement can be implemented using optical switches and linear optics as we detail in Appendix~\ref{app:xmeas}). Successfully detecting the photonic qudit in any of the Fourier basis states will herald the entangling operation and prepare the two qubit registers in a state 
\begin{equation}\label{eq:final_state}
\frac{1}{2^{m/2}}\sum_{l=0}^{2^m-1}\ket{1_l}_A\ket{1_l}_B=\frac{1}{2^{m/2}}(\ket{0}_A\ket{0}_B+\ket{1}_A\ket{1}_B)^{\otimes m} 
\end{equation}
up to single-qubit phase corrections dependent on the measurement outcome (see details in Appendix~\ref{app:xmeas})). Thus, Alice and Bob can create $m$ entangled qubit pairs by only transmitting a single photonic qudit between them.

Importantly, the multiple entangled pairs are created simultaneously, and the minimal required coherence time of the qubits in the registers only depends on the duration of a single entanglement attempt. For distant registers in an extended quantum network, this will be determined by the signaling time between the registers. Furthermore, assuming an overall transmission probability of $\eta$ between the two qubit registers, the entanglement generation rate of the protocol scales as $\eta$ since only one single photon is transmitted. 

The above protocol should be compared to the standard qubit approach where entanglement between two $m$-qubit registers would be attempted in parallel (see Fig.~\ref{fig:1}(b)). In the qubit approach, entanglement between a photonic qubit and a memory qubit would be generated separately for each of the memory qubits in the register. For applications that require the presence of all entangled pairs before execution such as the teleportation of a logical qubit composed of multiple physical qubits~\cite{Fowler2010,Munro2015,Muralidharan2016} or entanglement purification protocols~\cite{Bennett1996,Deutsch1996}, successful pairs need to be stored while waiting for the remaining pairs to succeed. Otherwise, simultaneous success in all qubit links would be necessary, leading to a success probability scaling as $\eta^m$ and thus causing an impractical rate for entanglement generation. Storing all successful pairs leads to a necessary coherence time of the qubits that scales as $1/\eta$ while the rate would scale as $\sim(2/3)^{\log(m)}\eta$ for $\eta\ll1$~\cite{Bernardes2011}. Compared with the qubit protocol, our qudit protocol both removes the unfavorable scaling of the memory time with $\eta$ ($t_0 \ \ {\rm v.s.}\ \ t_0/\eta$), and offers a rate that is more robust to transmission loss in terms of scaling with $m$ ($\eta \ \ {\rm v.s.}\ \ (2/3)^{\log(m)}\eta$). However, we do note that the local photon loss at the nodes can be different for the qubit and qudit approach depending on the specific implementation, which we will discuss in the following section.

\section{Implementation} \label{secIII}

\begin{figure*}[tp]
    \centering
    \includegraphics[width = \linewidth]{ 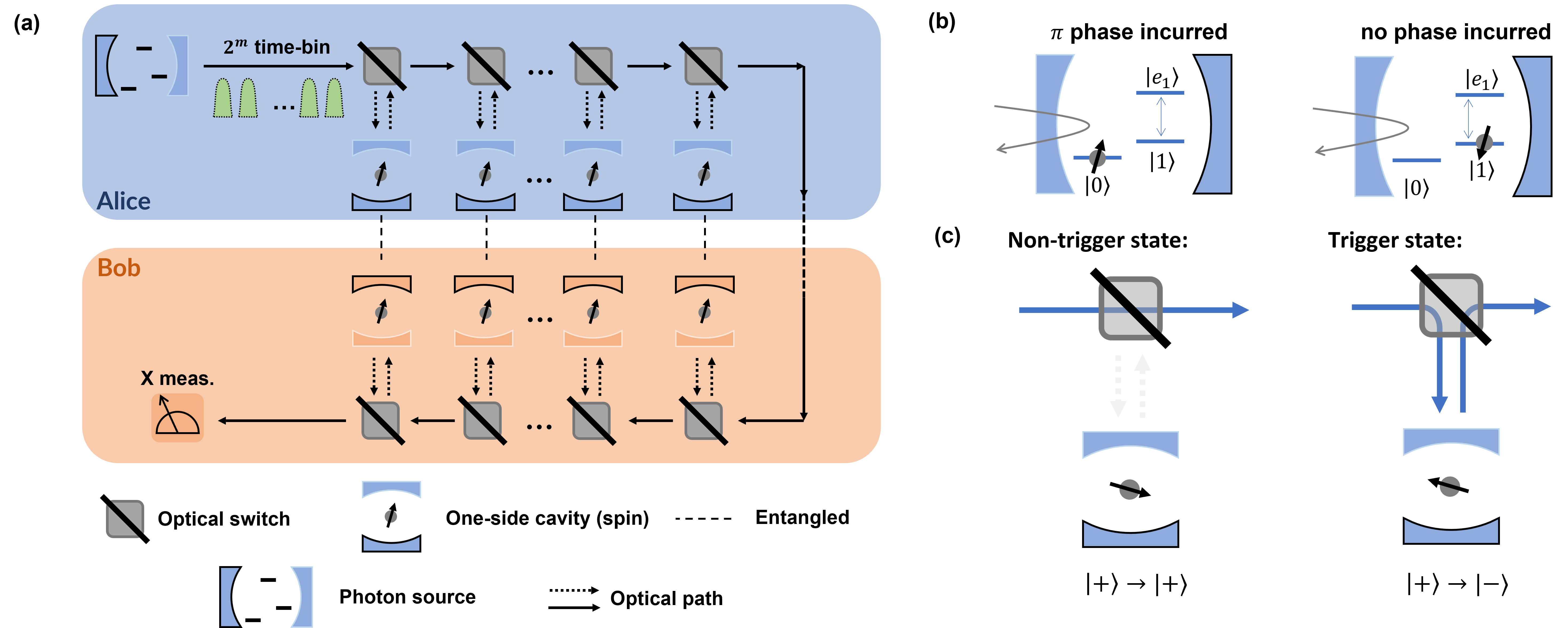}
    \caption{(a) A possible implementation of the proposed protocol using time-bin encoded photons and quantum emitters coupled to one-sided optical cavities. A pulsed cavity-assisted Raman scheme is used to generate the photonic time-bin qudit. Optical switches will route the photon to interact with different spin-cavity systems dependent on the binary encoding of the time-bin. The scattering of a photon from the one-side spin-cavity system implements a spin flip resulting in spin-photon entanglement. Interacting with both qubit registers (Alice and Bob) consecutively generates spin-spin entanglement, and a final X-basis measurement of the photonic qudit heralds the successful generation of multiple entangled pairs across the two registers. (b) Photon-spin CZ gates. A resonant photon will be reflected (gray arrow) from the cavity-spin system with (without) a $\pi$ phase shift if the qubit is in the uncoupled (coupled) state $\ket{0}$ ($\ket{1}$). This gate is equivalent to a photon mediated CNOT gate if Hadamard gates are performed on the spin qubit before and after the interaction.  (c) Sketch of the underlying mechanism behind the interaction scheme between the time-bin qudit and the qubit register. Consider a photon in time-bin $l$ with binary representation of $l$ being $[l_{m-1}...l_{k}...l_1 l_0]_{2}$. If $l_i=0$, the $i$'th optical switch will let the photon pass through without interaction with the $i$'th cavity-spin system. If $l_i=1$, the optical switch will instead route the photon to the cavity and flip the spin state through the interaction pictured in (b).}
    \label{fig:3}
\end{figure*}

 So far, we have only considered the high-level outline of our proposed protocol. Arguably, we have assumed a number of operations, which might be experimentally more demanding than standard spin-photon entanglement operations. Furthermore, the entangling operation between the photonic qudit and multiple memory-qubits can lead to correlated errors across the entangled qubit pairs that would not be present for parallel qubit approaches. 

To investigate the impact of these effects, we consider a specific implementation of our scheme based on single quantum emitters such as neutral atoms~\cite{Reiserer2014} or diamond defect centers~\cite{Bhaskar2020} coupled to optical resonators, which has already demonstrated some of the key functionality required for our scheme. In particular, we will exploit the photon-induced atomic phase gates based on single-sided cavities with a strongly coupled emitter~\cite{Duan2004}. We will refer to such systems as a spin-cavity system. The same system can also be used to generate high-dimensional photonic time-bin qudits by means of a pulsed, cavity-assisted Raman scheme~\cite{Knall2022}.

A sketch of the proposed implementation is shown in Fig.~\ref{fig:3}(a). In the first step, a photonic qudit is generated by a pulsed driving of a cavity-assisted Raman transition. Control of the driving power allows to tailor the amplitudes in the qudit state. This is crucial for the specific implementation considered here since the photon will experience different loss depending on which time-bin it is emitted in due to e.g. non-perfect interaction with a different number of spin-cavity systems in the spin-photon entangling step. For an initially even amplitude state (see Eq.~(\ref{eq:equation1})), this would decrease the fidelity of the entangled pairs. However, we can compensate the uneven loss by generating a qudit state with uneven amplitudes in such a way that the time-bin experiencing the most loss initially has the highest amplitude. This allows to move the effect from decreasing the fidelity to a modest decrease in rate. 

Dominant imperfections in the qudit generation step amount to finite spin coherence time of the emitter, imperfect pulse shaping of the driving laser, spontaneous emission from the excited state, and general photon loss (e.g. from absorption/material scattering). The latter will simply decrease the rate of the protocol given that no photon will be detected in the heralding step. The other imperfections will in general lead to an effective dephasing of the photonic qudit state due to leak of information to the environment about the emission time. Furthermore, imperfect driving may also lead to errors in the amplitude shaping of the qudit state. We refer to Appendix~\ref{app:ps} for a detailed optical model of the photon generation step.  

In the second step of the protocol, optical switches are used to route the photon to the spin-cavity systems as dictated the binary encoding of the time-bins (see Fig.~\ref{fig:3}(b)-(c)). We model the imperfections of the switches as consisting of both general loss and wrong switching (see Appendix~\ref{app:switch} for details). The spin qubits are initialized in the state $\ket{+}=(\ket{0}+\ket{1})/\sqrt{2}$ and ideally only the state $\ket{1}$ is coupled to an excited state, $\ket{e_1}$, by the cavity field. However, for systems such as SiV defect centers and quantum dots, the state $\ket{0}$ would also be coupled off-resonantly by the cavity field to another excited state, $\ket{e_0}$ (not shown in Fig.~\ref{fig:3}(b)). However, by tuning the frequency of the incoming photon and the cavity wrt. to the optical spin transitions, it is possible to realize a high-fidelity controlled phase gate where an incoming photon will flip the atomic state from $\ket{+}\to\ket{-}$. The combination of the switches and the spin-photon gates, followed by Hadamard gates on the spins, leads to the generation of the photon-spin entangled state in Eq.~(\ref{eq:spinqudit1}) in the ideal case.

\begin{figure*}[tp]
     \centering
     \includegraphics[width = \linewidth]{ 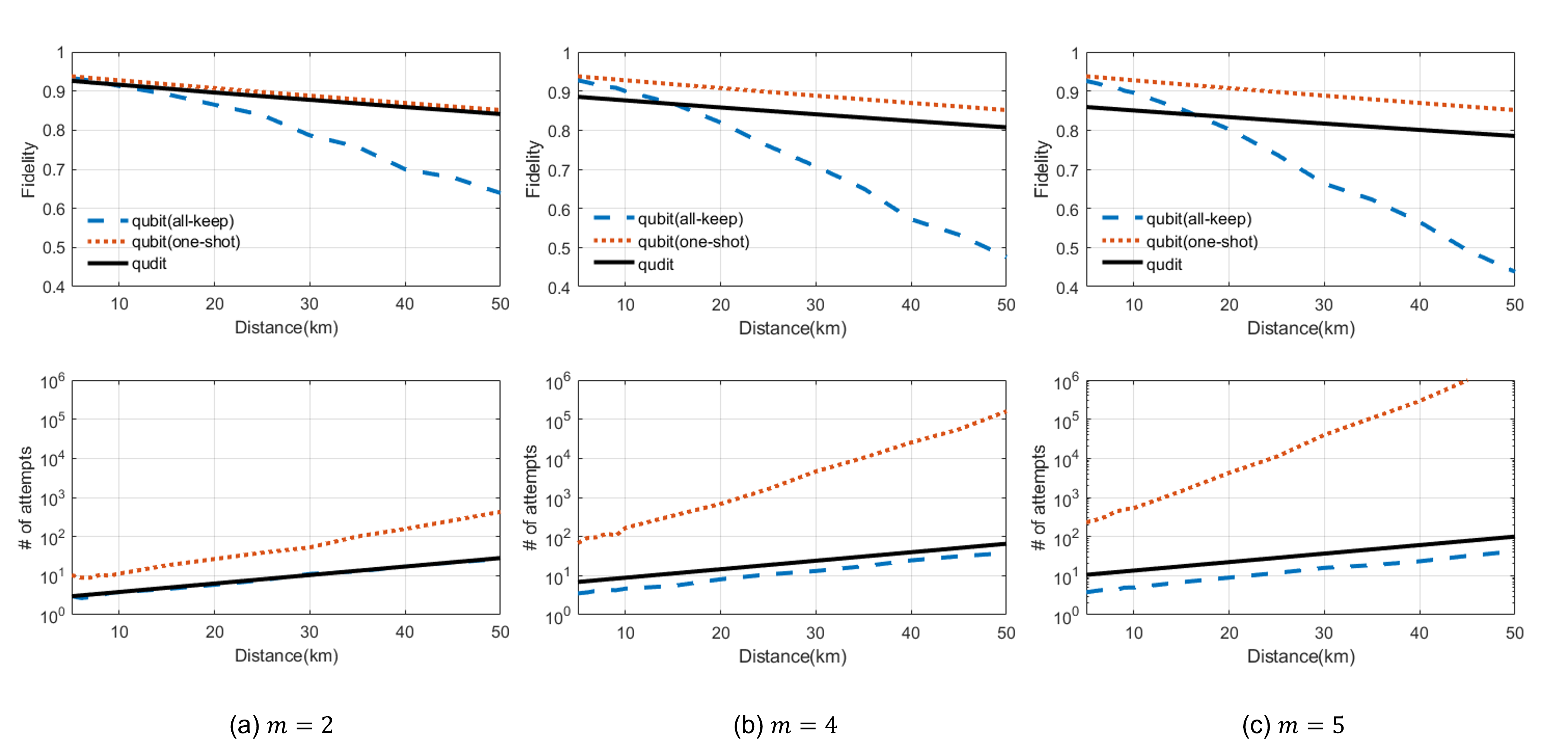}
    \caption{Performance comparison between the qudit protocol and the comparative qubit protocol for the generation of (a) $m=2$, (b) $m=4$ and (c) $m=5$ entangled pairs. The performance metrics are the average fidelity of the entangled pairs and the average number of attempts needed until the successful generation of all pairs. The fidelity is calculated as the average fidelity per Bell pair when tracing over the other spin qubits. For the qubit protocols, "all-keep" means the practical strategy where successful pairs are stored until all paris have been generated and "one-shot" means the strategy  where all entangled pairs are required to be generated simultaneously.  We have assumed that the time of local operations is negligible such that the rates of both the qudit and qubit protocols are set by the signaling time between the two distant registers. The simultaneous attempt of all links in the qubit protocol is counted as one attempt.}
    \label{fig:simulation}
\end{figure*}

The gate will ultimately be limited by the optical splitting between the $\ket{0}\leftrightarrow\ket{e_0}$ and $\ket{1}\leftrightarrow\ket{e_1}$ transitions, finite cooperativity of the emitter-cavity system, spectral width of the photon, and cavity loss assuming that the emitters and cavities can be tuned into the specific resonance conditions~\cite{Bradac2019}. In Appendix~\ref{app:gate} we give a detailed quantum optical model of the spin-photon interaction encompassing these imperfections.

After transmission to the second register, a similar system of optical switches and spin-cavity systems is used to create the state in Eq.~(\ref{eq:spinqudit2}) in the ideal case. As detailed above, the protocol is heralded by the generalized X-basis measurement of the qudit state. This can be implemented using optical switches and linear optics as detailed in Appendix~\ref{app:xmeas}. In particular, the switches are used to convert the time-bin encoding into a spatial encoding, where delay lines are used to ensure temporal coincidence. The quantum Fourier transform can then be implemented by means of a beam splitter circuit~\cite{Barak2007}, and the photon is finally detected with single photon detectors. 

Notably, the physical resources of the detection step such as number of switches, beam splitters and detectors scale linearly with the dimension of the photonic qudit. This can, in principle, be circumvented with probabilistic implementations using linear optics where beam splitters can be used to interfere multiple time-bins using a single fiber loop or interfere the  qudit photon with local single photons to erase the arrival time as outlined in Ref.~\cite{Khabiboulline2019pra}. Alternatively, deterministic approaches where a Raman-based absorption scheme in a three level emitter is attempted for each time-bin followed by atomic detection as described in Ref.~\cite{Khabiboulline2019} can be used to make the physical resources independent of the qudit dimension.

In the detection step, we model imperfect switches as in the entangling steps of the protocol (see Appendix~\ref{app:switch} for details). Notably, both loss and wrong switching lead to detectable errors. The former is due to the absence of a detection and the latter is due to a wrongly timed detection. Furthermore, we model loss in the delay lines which leads to a decrease in rate and not fidelity when balanced correctly in the qudit generation step. Finally, we model phase fluctuations in the beam splitters and optical paths as a general dephasing channel on the qudit state, where the dephasing parameter scales with the dimension of the system $m$. We note that the general dephasing channel can also incorporate other phase errors from e.g. general phase instabilities in optical paths   

\section{Simulation result}
We numerically simulate the simultaneous generation of multiple entangled pairs using our protocol and benchmark it against the parallel qubit approach using similar systems for increasing distance between the two nodes. As shown in Fig.~\ref{fig:simulation}, the qudit protocol quickly outperforms the qubit protocol (all-keep) for high-fidelity entanglement generation as the distance and hence transmission loss and signaling time increases. In general, we see that the qudit protocol has the same scaling in fidelity as the one-shot qubit protocol due to similar coherence time requirements while being as efficient as the all-keep qubit protocol. 

In our simulations, we have assumed the spin qubits to be subject to pure dephasing with a dephasing time of  $T_\phi=5{\rm ms}$ and a finite-temperature amplitude damping channel with decoherence time of $T_1=10{\rm ms}$~\cite{Nguyen2019,Tamara2021}. In the qudit generation, we have assumed to be limited by amplitude and phase fluctuations in the laser modelled as Gaussian noise with variances $\sigma_a^2 = \sigma_p^2 = 0.01$. We have assumed 10\% loss for the photonic switches and 1\% switching error, 5\% loss in the input/output coupling from the cavity, a cooperativity of C=100 for the spin-cavity system and negligible coupling of the $\ket{0}$ state. We note that while our qudit protocol is more robust to transmission loss than the qubit approach, it is less robust to local loss. Letting $\eta_{0}$ be the efficiency of transmission of the photon when interacting with a single switch and cavity, the total transmission of the qudit protocol will scale $\eta_0^m$. Nonetheless, we see from Fig.~\ref{fig:simulation} that even for $\eta_0\sim 86\%$ and $m=5$, the qudit protocol significantly outperforms the qubit protocol for high-fidelity entanglement generation for distances $\gtrsim 20$ km due to the much relaxed requirements on the qubit coherence times.

We also include a general qudit dephasing channel modelled as random phase shifts of the time-bin states following a Gaussian distribution with zero mean and variance $\sigma=0.1 m$. We assume the scaling with $m$ to account for the additional complexity of the interferometer and phase instability in optical paths (See Appendix \ref{app:xmeas} for details). This affects the fidelity of the qudit protocol resulting in a worse performance as $m$ increases as visible from Fig.~\ref{fig:simulation}. We note that this increased dephasing, however, is somewhat arbitrary given that it models a technical noise and not a fundamental one. 

In our simulations, we have assumed that the time of the local operations is negligible compared to the signaling time between the two stations, which is a valid assumption for extended networks. Node distances for realistic quantum networks are usually considered to be tens of kilometres corresponding to signaling times around 0.1-1 ms~\cite{Borregaard2015pra,Uphoff2016,Huie2021}. This is, in general, much longer than the time of local operations, which sets the duration of a photonic time-bin. Purcell enhanced emission and fast switching can be envisioned leading to time-bins in the regime of 10 nanoseconds~\cite{borregaard2020} and the time of local operations thus only become relevant for qudit dimensions of $\gtrsim10^3$ corresponding to $m \gtrsim 10$. We note that in practice $m$ will, however, most likely be limited  to $<10$ due to the growing complexity of the implementation such as the generation of the many time-bins as well as calibration and phase stabilization of the optical path. Thus, the time of one entanglement generation attempt is $T_{\text{gen}}\sim L/c$, where $L$ is the distance between the stations and $c$ is the speed of light in an optical fiber. For fiber propagation, the transmission efficiency of a photon between the stations will be $\eta_{\text{f}}=e^{-L/L_{\text{att}}}$, where $L_{\text{att}}$ is the attentuation length of the fiber. In our simulations, we have assumed $L_{\text{att}}=20$ km corresponding to fiber loss for telecom light. We note that for systems such as diamond defect centers and neutral atoms, frequency conversion from optical to telecom would be required.  

It is clear that if $T_{\text{coh}}\gg T_{\text{gen}}e^{L/L_{\text{att}}}$, the qudit protocol does not offer much advantage compared to the standard qubit approaches. We note, however, that this requires the coherence time to increase exponentially with the distance. This exponential dependence can be circumvented either with the use of quantum repeaters~\cite{Munro2015} or by employing multiplexing~\cite{Collins2007} where the number of qubit memories far exceeds the number of desired Bell pairs. In the latter case, the required number of memories will also increase with the transmission loss. Both approaches can, in principle, be combined with the proposed qudit protocol for entanglement generation to further alleviate memory requirements and decrease resources. In particular, both for entanglement purification in 1st generation repeaters and in the operation of 2nd generation repeaters, multiple and high-fidelity Bell pairs are required between neighboring stations~\cite{Munro2015,Muralidharan2016}. Employing the qudit protocol to alleviate the memory requirements can lead to longer distances between links in the repeater thereby also reducing the overall resources for long-distance entanglement distribution. This, in particular, may also be beneficial for highly-lossy links in satellite-based entanglement distribution~\cite{Juan2017}. 

\section{Conclusion and Discussion}

We have proposed a novel entanglement distribution protocol using photonic qudits, which significantly alleviates the requirement for coherence time compared with the canonical protocol and provides benefits for near-term quantum networks. 

 We note that while we discussed a specific implementation based on scattering of single-sided cavities and optical switches, a number of alternative implementations can be imagined that align with current experimental efforts on quantum repeaters with solid state spins. The optical switches can be replaced by periodic spin rotations in between the time-bins as demonstrated in Ref.~\cite{Bhaskar2020} with SiV centers and the multiple cavity systems can be replaced with a single emitter coupled to a nearby qubit register as demonstrated in Ref.~\cite{Nguyen2019}. If controlled rotations of the emitter spin system dependent on the state of the qubit register can be performed in between the time-bins, the same qudit-qubit entanglement can be generated. Our work thus integrates with current experimental state-of-the-art for quantum communication.  
 
 An alternative implementation as an emission-based scheme where the qubit register controls whether the emitter is in an optically bright or dark state similar to entangling schemes demonstrated with NV centers~\cite{Hensen2015,Pompili2021} can also be envisioned. Performing a generalized Bell measurement at a middle station could swap the qubit-qudit entanglement to qubit-qubit entanglement.
  
For single emitter systems such as defect centers in diamond, coupled to nearby nuclear qubit registers, the coherence time of the nuclear qubits is often severely reduced when operations on the emitter are performed. Generation of multiple Bell pairs with such a system requires storage of already generated Bell pairs in the qubit register while attempting another Bell pair with the emitter. As a result, the stored Bell pairs will quickly decohere, limiting the performance of the qubit protocol~\cite{Kalb2018}. The qudit approach circumvents this since all Bell pairs are generated simultaneously. We thus believe that our protocol is particularly relevant for enhancing the performance of current quantum network systems based on diamond defect centers such as NV and SiV centers.


\begin{acknowledgements}
We thank Y. Blanter and R. Hanson for illuminating discussions and useful comments. Y. Zheng thanks the support from TU Delft Excellence Scholarship.  This work was supported by the NWO Gravitation Program Quantum Software Consortium. 
\end{acknowledgements}

\appendix

\section{Photonic qudit source}\label{app:ps}
\begin{figure}[t]
    \centering
    \includegraphics[width = 0.8\linewidth]{ 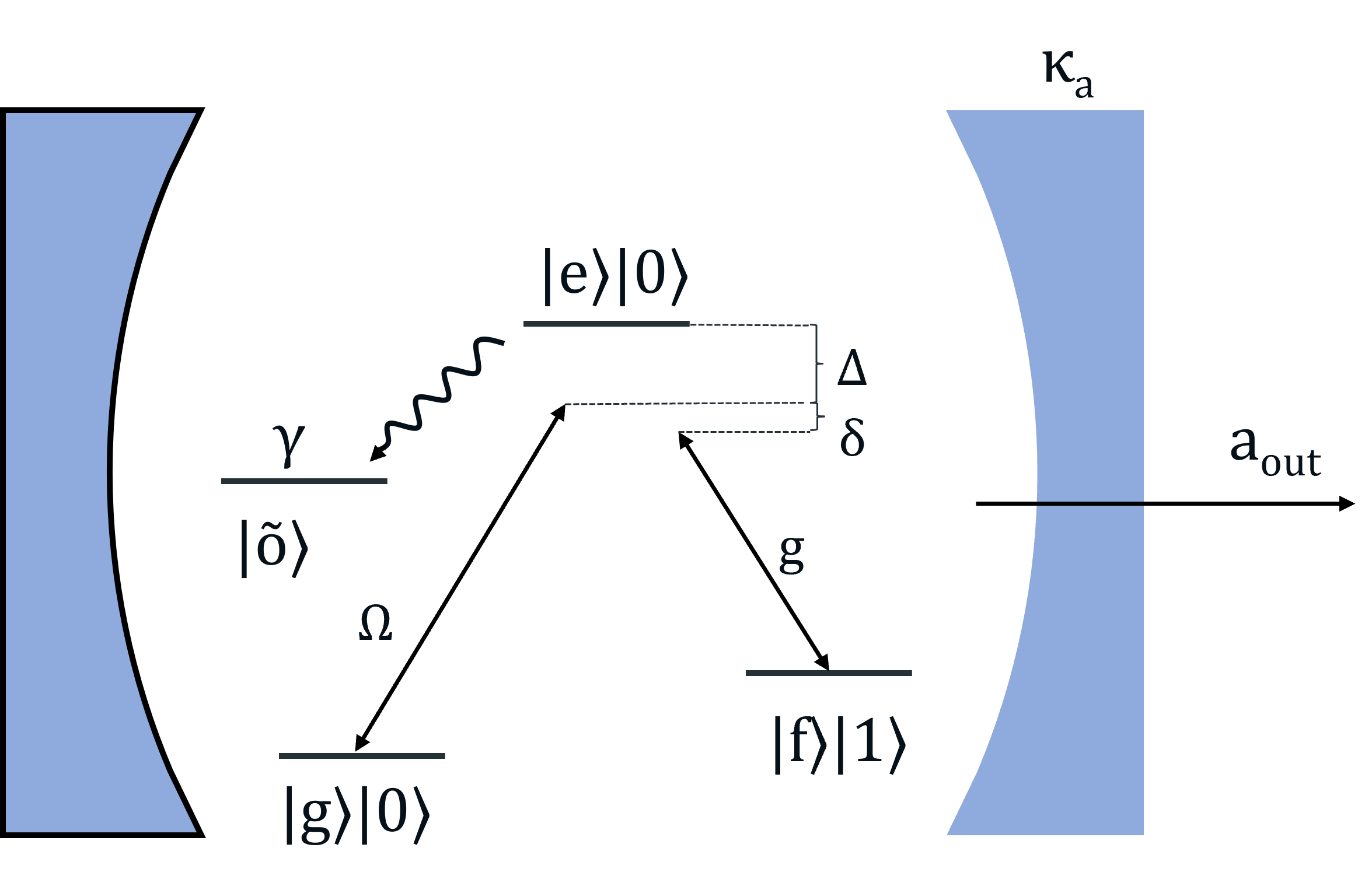}
    \caption{The level structure of the photonic qudit source.}
    \label{fig:emitter}
\end{figure}

The photonic qudit can be emitted from a three-level $\Lambda$ system with two ground states $\ket{g}$ and $\ket{f}$ and an excited state $\ket{e}$ (see Fig.~\ref{fig:emitter}). The transition $\ket{g}\to\ket{e}$ is driven by a laser pulse while a single mode cavity field couples the transition $\ket{e}\to\ket{f}$. The Hamiltonian of the system in a suitable rotating frame and after making the rotating wave approximation can be described as ($\hbar=1$)
\begin{equation} \label{eq: hjc}
    \hat{H} = \Delta \ket{e}\bra{e} +(\Omega \ket{e} \bra{g}+H.C.)+ (g\ket{e}\bra{f}c + H.C.)
\end{equation}
where, $\Omega$ describes the laser coupling, $\Delta$ is the detuning between the laser frequency and the $\ket{g} \xrightarrow{} \ket{e}$ transition and $g$ is the coupling between the spin and the cavity field.

We describe spontaneous emission from the excited state and the decay of the cavity field with Lindblad jump operators
\begin{equation}
\begin{aligned}
\hat{L}_{\gamma_g} &= \sqrt{\gamma_g} \ket{g}\bra{e}\\
\hat{L}_{\gamma_f} &= \sqrt{\gamma_f} \ket{f}\bra{e}\\
\hat{L}_\kappa &= \sqrt{\kappa}  \hat{c}
\end{aligned}
\end{equation}
where $\gamma_g$ is the rate of decay from $\ket{e}$ to $\ket{g}$, $\gamma_f$ is the rate of decay from $\ket{e}$ to $\ket{f}$ and $\kappa$ is the cavity decay rate. We neglect pure dephasing of the excited state which anyway will have a negligible effect for large detuning. Furthermore, we neglect decay of the ground states assuming that this is negligible on the time-scale of the photon generation.  
Adopting the stochastic wave function picture~\cite{Dalibard1992}, the system evolves according to the effective Hamiltonian  
\begin{equation}\label{eq: sh}
\begin{aligned}
\hat{H}_{\text{eff}} = \Delta \ket{e}\bra{e} + \delta \hat{c }^\dagger\hat{c} + (g\ket{e}\bra{f}c + H.C.)\\ + (\Omega \ket{e} \bra{g}+H.C.)
- \frac{i}{2}\left(\gamma\ket{e}\bra{e}+\kappa \hat{c}^\dagger\hat{c} \right)
\end{aligned}
\end{equation}
where, $\gamma = \gamma_g + \gamma_f$ in the absence of decay.  

We can express the wave function of the system under the evolution of $\hat{H}_{\text{eff}}$ in the basis of $\ket{g,0}$, $\ket{e,0}$ and $\ket{f,1}$ where $\ket{g,0}$ ($\ket{f,1}$) denotes the emitter in state $\ket{g}$ ($\ket{f}$) and no (one) cavity photon. The wave function can thus be expressed as 
\begin{equation}
\ket{\phi(t)} = c_0(t)\ket{g,0} + c_1(t)\ket{e,0} + c_2(t)\ket{f,1},
\end{equation}
and we obtain the equations of motion from the Sch\"{o}dinger equation $i\frac{\partial\ket{\phi}}{\partial t}=\hat{H}_{\text{eff}}\ket{\phi}$. Note that $\hat{H}_{\text{eff}}$ is not Hermitian, which results in a decrease of the norm of $\ket{\phi}$. This reflects that the evolution is in the absence of decay and $\braket{\phi(t)}{\phi(t)}$ is the probability that no decay has happened.  
The Schrodinger equation gives us a set of differential equations in the time-dependent coefficients,

\begin{equation}\label{eq: diff1}
\begin{split}
i \dot{c}_0 &= c_1 \Omega ^*    \\
i \dot{c}_1 &= c_0 \Omega + c_1 \left(\Delta - \frac{i}{2}\gamma\right) +c_2 g\\
i \dot{c}_2 &= c_1g^*+c_2\delta - \frac{i}{2}\kappa c_2
\end{split}    
\end{equation}
When solving these differential equations we make a few realistic assumptions. First, we assume that the laser drives the system adiabatically such that $\Omega \ll \Delta,\gamma$ and that we are in the bad cavity regime where $\gamma \ll g \ll \kappa$ since we are interested in getting a photon out of the cavity. This is also the relevant regime for nanophotonic cavities~\cite{Bhaskar2020,Bradac2019}. Secondly, we assume that the two-photon detuning is zero, i.e. $\delta = 0$. Furthermore, we will assume that the driving can be approximated as a square pulse of duration $T_1$. Under these assumptions we can adiabatically eliminate $\dot{c}_1$ and $\dot{c}_2$ and solve the resulting differential equation for $c_2(t)$. For $0<t<T_1$, we find that 
\begin{equation}
    c_2(t) = \frac{2i\Omega g^* \left(\Delta + \frac{i\gamma}{2}(1+4C)\right)}{\kappa\left(\Delta^2 + \frac{\gamma ^2}{4}(1+4C)^2\right)}e^{bt} c_0(0)
\end{equation}
where, $c_0(0)$ is the population in the $\ket{g,0}$ state at the start of the laser pulse and $b$ is
\begin{equation*}
b =     \frac{i|\Omega|^2 (\Delta + \frac{i\gamma}{2}(1+4C))}{\Delta^2 + \frac{\gamma ^2}{4}(1+4C)^2},
\end{equation*}
where we have defined the cooperativity $C=\abs{g}^2/\kappa\gamma$.

After the pulse, any population in the excited state will decay by the emission of a cavity photon to the $\ket{f,1}$ state. Solving the differential equations for the evolution of $c_2(t)$ for $t>T_1$ gives us
\begin{equation}
\begin{split}
    c_2(t) = \frac{2ig^*\Omega (\Delta + \frac{i\gamma}{2}(1+4C))}{\kappa\left(\Delta^2 + \frac{\gamma ^2}{4}(1+4C)^2\right)} \times \\
    e^{bT_1} e^{-i\left(\Delta - \frac{i\gamma}{2}(1+4C) \right)(t-T_1)}c_0(0)
    \end{split}
\end{equation}
We can observe that after the driving is stopped, the population in the excited state decays to the $\ket{f,1}$ state exponentially with the Purcell enhanced rate $\gamma(1+4C)$. We note that the expression of $c_2(t)$ for a square pulse depends on the initial population of the $\ket{g,0}$ state, the amplitude of the drive and the duration of the drive. In order to make time-bins indistinguishable, we will have to change the amplitude of the driving from time-bin to time-bin in order to keep the duration of the pulse fixed.

The mode of the output time-bin, $v(t)$, is found using the standard input-output relations of the cavity giving 
\begin{equation}\label{eq: output_mode}
\begin{aligned}
    v(t<T_1) &= \frac{2i\Omega g^* \left(\Delta + \frac{i\gamma}{2}(1+4C)\right)}{\sqrt{\kappa}\left(\Delta^2 + \frac{\gamma ^2}{4}(1+4C)^2\right)}e^{bt} c_0(0)\\
    v(t>T_1) &= \frac{2ig^*\Omega (\Delta + \frac{i\gamma}{2}(1+4C))}{\sqrt{\kappa}\left(\Delta^2 + \frac{\gamma ^2}{4}(1+4C)^2\right)}\times\\ 
    & e^{bT_1} e^{-i\left(\Delta - \frac{i\gamma}{2}(1+4C) \right)(t-T_1)}c_0(0).
\end{aligned}
\end{equation}
We see that the phase of the photon and the amplitude depends on the phase and strength of the laser pulse. Phase and amplitude fluctuations of the driving laser will thus directly affect the phase and amplitude of the specific time-bin in the qudit state. 

From Eq.~(\ref{eq: output_mode}), the probability of any spontaneous emission during the drive can be found as 
\begin{equation}
\begin{aligned}
P_{\gamma} &=1-  \frac{4C}{1+4C}\left(1- e^{2AT_1}\right) \\&-\frac{4C}{1+4C}\left(\frac{|\Omega|^2e^{2AT_1}}{\Delta^2 + \frac{\gamma^2}{4}(1+4C)^2}\right)\\
&\approx 1-\frac{4C}{1+4C}\left(1- e^{2AT_1}\right)
\end{aligned}
\end{equation}
where,
\begin{equation*}
    A = -\frac{|\Omega|^2\gamma(1+4C)}{2\left(\Delta^2 + \frac{\gamma^2(1+4C)^2}{4}\right)}.
\end{equation*}
With probability $\frac{\gamma_f}{\gamma}P_{\gamma}$ such spontaneous emission would result in no cavity photon being emitted since the emitter will be trapped in the $\ket{f}$ state. Thus, this effectively amounts to photon loss. With probability $\frac{\gamma_g}{\gamma}P_{\gamma}$, the spontaneous emission would bring the emitter back to the state $\ket{g}$ from which a photon could subsequently be emitted in a later time-bin. This decay leaks information to the environment that the photon had not been emitted in any prior time-bins and thus effectively acts as dephasing of the photonic qudit. We note, however, that both processes are suppressed with the cooperativity as $1/C$.  

In the high cooperativity limit, the main dephasing of the photonic qudit arguably originates from laser noise which we model as Gaussian noise. We thus decribe the generated photonic qudit as 

\begin{equation} \label{eq:quditgen}
    \ket{\psi}_{\text{noisy}} = \frac{1}{2^{m/2}\sqrt{\mathcal{N}}}\sum^{2^m-1}_{i=0}(1+\alpha_i)e^{i\theta_i}\ket{i}
\end{equation}

Where $\mathcal{N}=\sum^{2^m-1}_{i=0}(1+\alpha_i)$ is a normalization factor and $\alpha_i$ models the amplitude fluctuation following a Gaussian distribution $ N(0,\sigma_a^2)$, and $\theta_i$ is the phase fluctuation following a Gaussian distribution $ N(o, \sigma_p^2)$. In the simulation, we assumed $\sigma_a = \sigma_p = 0.1$. 

\section{Spin-photon entangling gate}\label{app:gate}
\begin{figure}[t]
    \centering
    \includegraphics[width =.8 \linewidth]{ 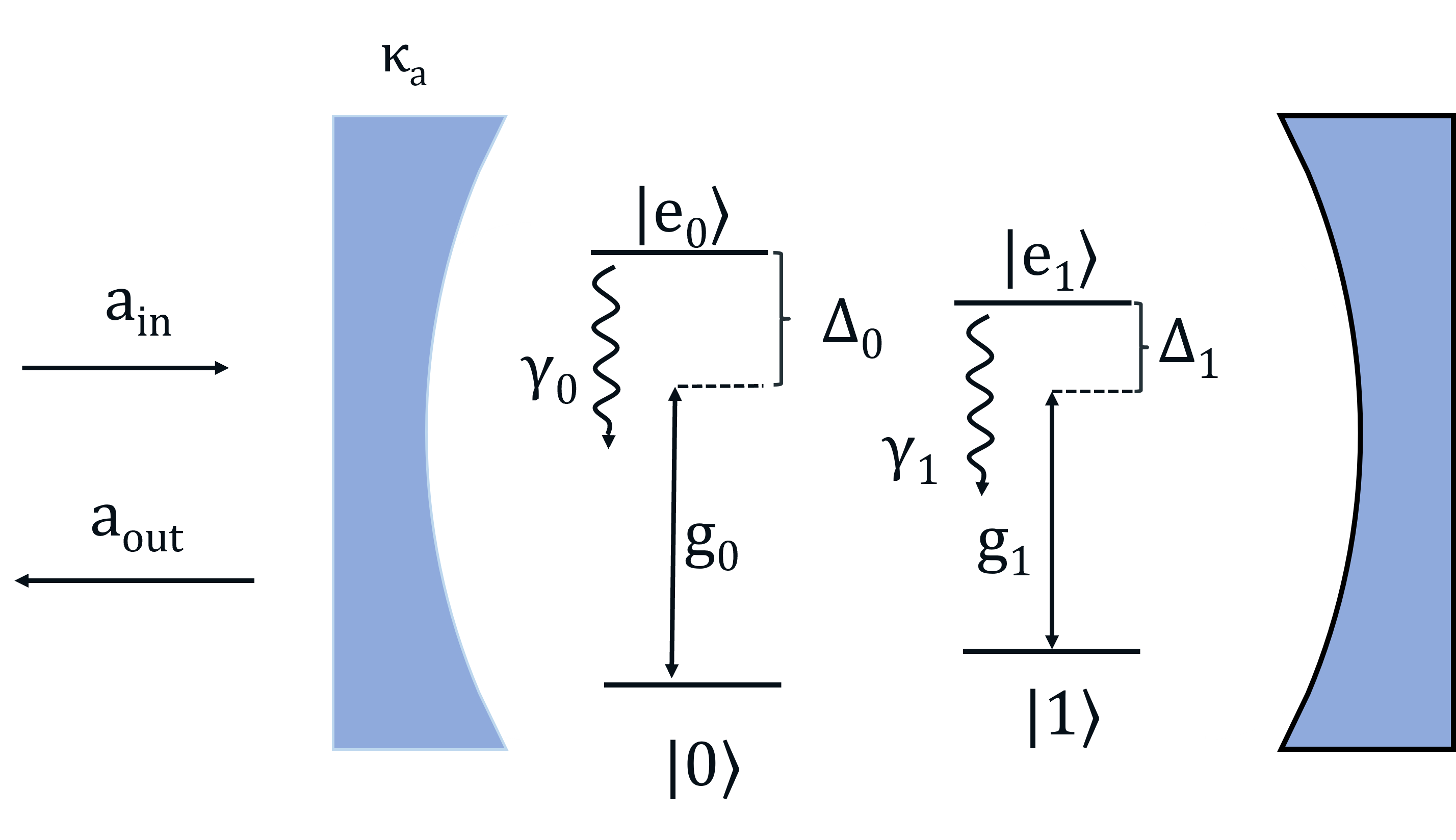}
    \caption{The four-level system for spin-photon entanglement.}
    \label{fig:gate}
\end{figure}

In this section we describe the cavity mediated spin-photon CZ gates. For realistic systems such as diamond vacancy centers or quantum dots~\cite{Tiurev2021}, we need to consider a 4-level system with two ground states, $\ket{0}$ and $\ket{1}$, and their respective excited states, $\ket{e_0}$ and $\ket{e_1}$ (see Fig.~\ref{fig:gate}). We assume that the cavity field couples both ground states $\ket{0}$ and $\ket{1}$ to excited states $\ket{e_0}$ and $\ket{e_1}$, respectively. The Hamiltonian of the system in the rotating frame w.r.t. the cavity mode is
\begin{equation}
\begin{aligned}
\hat{H} = \Delta_0\ket{e_0}\bra{e_0}+ \Delta_1\ket{e_1}\bra{e_1} + \left(g_0\ket{e_0}\bra{0}\hat{c}+H.C. \right)\\
+ \left(g_1\ket{e_1}\bra{1}\hat{c}+H.C., \right)
\end{aligned}
\end{equation}
where $\Delta_0$ and $\Delta_1$ are the detunings between the cavity mode and the transitions $\ket{0} \leftrightarrow\ket{e_0}$ and $\ket{1} \leftrightarrow \ket{e_1}$ respectively and $g_1$ ($g_0$) is the cavity coupling of the $\ket{1}\leftrightarrow\ket{e_1}$ ($\ket{0}\leftrightarrow\ket{e_0}$) transition. We model spontaneous emission with Lindblad Jump operators.
\begin{equation}
\begin{split}
L_0 &= \sqrt{\gamma_0} \ket{0}\bra{e_0}\\
L_1 &= \sqrt{\gamma_1} \ket{1}\bra{e_1}\\
\end{split}
\end{equation}
where, $\gamma_0$ and $\gamma_1$ are rates of decay of excited states. Note that we have neglected spontaneous emission on the cross transitions in the system ($\ket{e_0}\leftrightarrow\ket{1}$, $\ket{e_1}\leftrightarrow\ket{0}$) for simplicity. Any spontaneous emission would effectively be a loss of the incoming photon and thus would be heralded in the entanglement generation attempt. Including cross transitions would thus not result in any change in the analysis. We have also neglected pure dephasing of the excited state and dephasing or decay of the ground states. Pure dephasing would broaden the spectrum of a reflected photon, which, in principle, can be turned into an effective loss by frequency filtering. Furthermore, the coherence of the ground states is arguably much longer than the photon interaction time in order for the system to be relevant for long-distance entanglement generation.  
\
The input-output relations for the (one-sided) cavity for an incoming field $\hat{a}$  are,
\begin{equation} \label{eq:inputoutput}
\begin{split}
\hat{a}_{out} &= \hat{a}_{\text{in}} - \sqrt{\kappa_a}\hat{c}\\
\dot{\hat{c}} &= i[\hat{H}_{JC}, \hat{c}] - \frac{\kappa \hat{c}}{2} + \sqrt{\kappa_a} \hat{a}_{\text{in}}, 
\end{split}
\end{equation}
where $\kappa_a$ denotes the decay rate of the cavity field to the collected mode while $\kappa=\kappa_a+\kappa'$ is the total cavity decay rate where $\kappa'$ is the decay due to intra cavity loss. Following the approach of Ref.~\cite{Anders2003},  we solve the differential equation for $\hat{c}$ in the Fourier domain and make the assumption that the incoming field is weak, i.e.
\begin{equation}
\begin{aligned}
\hat{\sigma}_{j,z} &= \ket{j}\bra{j} - \ket{e_j}\bra{e_j}  \\ 
&\approx \ket{j}\bra{j} = \hat{P}_j 
\end{aligned}    
\end{equation}
where, $\hat{P}_j$ is the projector onto the ground state $\ket{j}$ at time $t=0$ i.e. at the start of the photon scattering. The solution for $\hat{c}(\omega)$ in the Fourier domain is,
\begin{equation}
\hat{c}(\omega) = \frac{\sqrt{\kappa_a}\hat{a}_{\text{in}}(\omega)}{\left( i\omega + \frac{\kappa}{2} +  \sum_{j=0}^{1} \frac{|g_j|^2\hat{P}_j}{i(\omega + \Delta _j)+\frac{\gamma_j}{2}}    \right)},
\end{equation}
where we have neglected any noise operators from cavity decay or spontaneous emission since these only lead to vacuum in the output i.e. an effective loss of the incoming photon. 

If we input this expression in the input-output relations (Eq.~(\ref{eq:inputoutput})), we get a relation between the incoming field and the outgoing field,
\begin{equation}
\begin{split}
\hat{a}_{out}(\omega) &= \left[ 1 -  \frac{{\kappa_a}}{ i\omega + \frac{\kappa}{2} +  \sum_{j=0}^{1} \frac{|g_j|^2\hat{P}_j}{i(\omega + \Delta _j)+\frac{\gamma_j}{2}}    } \right] \hat{a}_{in}(\omega)
\end{split}
\end{equation}
We can therefore define the reflection coefficients as,
\begin{equation}
\begin{aligned}
    \hat{r}(\omega) = \left[ 1 -  \frac{{\kappa_a}}{ -i\omega + \frac{\kappa}{2} +  \sum_{j=0}^{1} \frac{|g_j|^2\hat{P}_j}{-i(\omega + \Delta _j)+\frac{\gamma_j}{2}}    } \right]\\ 
\end{aligned}    
\end{equation}

We note that $\omega$ is the detuning between the incoming photon and the cavity resonance frequency. For a long driving pulse compared to the (Purcell enhanced) decay rate of the emitter, we can ensure that the incoming photon is narrow in frequency compared to both the cavity and emitter linewidth. We will therefore assume that we can treat the incoming photon as a delta function at $\omega$. The reflection coefficients $r_0$ and $r_1$ for the ground states $\ket{0}$ and $\ket{1}$ are,
\begin{equation}\label{r0r1}
\begin{aligned}
r_0 &= \left[ 1 -  \frac{{\kappa_a/\kappa}}{ \frac{-i\omega}{\kappa} + \frac{1}{2} +  \frac{C_0}{\frac{-i}{\gamma_0}( \omega + \Delta _0)+\frac{1}{2}}    } \right]\\
r_1 &= \left[ 1 -  \frac{{\kappa_a/\kappa}}{ \frac{-i\omega}{\kappa} + \frac{1}{2} +  \frac{C_1}{\frac{-i}{\gamma_1}( \omega + \Delta _1)+\frac{1}{2}}    } \right],
\end{aligned}
\end{equation}
where we have defined the Cooperativities $C_0=\abs{g_0}^2/\kappa\gamma_0$ and $C_1=\abs{g_1}^2/\kappa\gamma_1$. 

Let us assume that only a single time-bin ($\ket{l}$) comes in and interacts with spin in the $\ket{+}$ state. In this case, the photon reflects off the cavity with reflection coefficients described in eq. \ref{r0r1}, and the photon-spin state undergoes the following transformation.

\begin{equation}
    \frac{1}{\sqrt{2}}\big( \ket{l}_{ph}\ket{0}_{s} + \ket{l}_{ph}\ket{1}_{s}\big) \xrightarrow{}\frac{1}{\sqrt{2}}\ket{l}_{ph}\big( r_0\ket{0}_{s}+r_1\ket{1}_{s}\big)
\end{equation}
In the ideal case, $r_0$ and $r_1$ assume values -1 and 1 respectively. For these values, the spin state is flipped from $\ket{+}$ to $\ket{-}$ after interaction of the spin with a time-bin qubit. To get a perfect CZ gate we can tune the parameters of the spin-cavity system. For $C_0=0$, the original CZ gate of Ref.~\cite{Duan2004} is recovered. For $C_0\neq 0$, a high fidelity CZ gate can still be obtained by tuning the frequency of the incoming photon and the cavity. In the simulation, we assume $C_0/\Delta_0\approx 0$, $C_1 = 100$, and $\kappa_a/\kappa = 0.95 $.
\section{Optical switches}\label{app:switch}

We used two parameters to model realistic switches: 

\begin{enumerate}
    \item switch transmission $\eta_{sw}$: $(1-\eta_{sw})$ is the probability of loss when a photon passes through a switch. The loss is uniform for the photonic qudit regardless of its time-bin state. In our simulation, we assumed $\eta_{sw}=0.9$.
    \item wrong switching $e_{sw}$: This is the fraction of the light that leaks to the wrong output port rather than the desired output port. For example, if the switch is turned off and all input should come out from output port 1 (passing through), the real case could be 
    \begin{equation}
    \ket{{\rm In}}\xrightarrow{}\sqrt{1-e_{sw}}\ket{{\rm Out}}_1 + \sqrt{e_{sw}}\ket{{\rm Out}}_2
\end{equation}
A $e_{sw}$ portion of the input is leaked from output port 2 (interacting with the cavity). In our simulation, we assumed $e_{sw}=0.01$.
\end{enumerate}

\section{X measurement}\label{app:xmeas}

The generalized X basis for a qudit with dimension $N=2^m$ can be described as
\begin{equation}
    \ket{X_k} = \dfrac{1}{\sqrt{N}}\sum^{N-1}_{l=0}\omega^{kl}\ket{l},
\end{equation}
where $\omega=e^{i\pi/2^{m-1}}$. $\ket{X_k}(k=0,1,...,N-1)$ compose a set of complete orthogonal measurement basis. For the simplest case when $m=1$, we have
\begin{equation}
    \begin{cases}
    &    \ket{X_0} = \dfrac{1}{\sqrt{2}}(\ket{0}+\ket{1})\\
    &    \ket{X_1} = \dfrac{1}{\sqrt{2}}(\ket{0}-\ket{1}).
    \end{cases}
\end{equation}
This is exactly the X-basis for qubit measurement.  For $\ket{X_k}$, the corresponding measurement operator is given by
\begin{equation}
    M_k = \ket{X_k}\bra{X_k},
\end{equation}
and measurement in such a basis will project a state $\rho$ into
\begin{equation}
    \rho_{meas} = \dfrac{M_k\rho M_k^\dagger}{{\rm Tr}(M_k \rho)}.
\end{equation}
From Eq. \ref{eq:spinqudit2}, the collective state right before the X measurement is
\begin{equation}
    \ket{\Psi_2}=\frac{1}{2^{m/2}}\sum_{l=0}^{2^m-1}\ket{l}_{ph}\ket{\tilde{1}_l}_A\ket{\tilde{1}_l}_B.
\end{equation}
Let's express the state in the X photonic basis. Notice that
\begin{equation}
    \ket{l} = \frac{1}{\sqrt{N}}\sum^{N-1}_{k=0}\omega^{-kl}\ket{X_k}
\end{equation}
so we have
\begin{equation}
    \ket{\Psi_2}=\frac{1}{N}\sum_{l=0}^{N-1}\sum^{N-1}_{k=0}\omega^{-kl}\ket{X_k}_{ph}\ket{\tilde{1}_l}_A\ket{\tilde{1}_l}_B.
\end{equation}
When partially measuring the photonic state of $\rho = \ket{\Psi_2}\bra{\Psi_2}$(Eq. \ref{eq:spinqudit2}), we have 
\begin{equation}
    \rho_{meas, k} = \dfrac{(M_k\otimes I_{AB})\ket{\Psi_2}\bra{\Psi_2}(M_k \otimes I_{AB})}{{\rm Tr}(M_k\otimes I_{AB} \ket{\Psi_2}\bra{\Psi_2}) }.
\end{equation}
Assume the photonic measurement outcome corresponds to the basis $\ket{X_k}$, the spin state after photonic measurement is still a pure state which is given by
\begin{equation}\label{eq:after_meas}
    \ket{\Psi}_{meas, k}=\frac{1}{\sqrt{N}}\sum_{l=0}^{2^m-1}\omega^{-kl}\ket{\tilde{1}_l}_A\ket{\tilde{1}_l}_B.
\end{equation}
Using $\ket{\tilde{1}_l}=\ket{l_{m-1}}\ket{l_{m-2}}...\ket{l_1} \ket{l_0}$,
\begin{equation}
\begin{split}
     \ket{\Psi}_{meas, k}&=\frac{1}{\sqrt{N}}\sum_{l=0}^{2^m-1}\omega^{-kl}\ket{l_{m-1}}_{A_{m-1}}...\ket{l_1}_{A_1} \ket{l_0}_{A_0}\\
     &\otimes\ket{l_{m-1}}_{B_{m-1}}...\ket{l_1}_{B_1} \ket{l_0}_{B_0},   
\end{split}
\end{equation}
where the subscript $A_p$ stands for Alice's $p$'th spin qubit, and $B_p$ stands for Bob's $p$'th spin qubit. Reorder the sequence of these spin qubits, and we will arrive at
\begin{equation}\label{eq:xmeas-general}
\begin{split}
        \ket{\Psi}_{meas, k} = \frac{1}{\sqrt{N}} & (\ket{00}+\omega^{-k} \ket{11})_{A_0 B_0}\\
        &\otimes(\ket{00}+\omega^{-2k}\ket{11})_{A_1 B_1}\\
        &\otimes (\ket{00}+\omega^{-4k}\ket{11})_{A_2 B_2} \otimes...\\
        &\otimes (\ket{00}+\omega^{-2^{m-1}k}\ket{11})_{A_{m-1} B_{m-1}}.
\end{split}
\end{equation}
If the measurement outcome corresponds to $k=0$, Eq. \ref{eq:xmeas-general} will be reduced to the tensor product of multiple entangled pairs (Eq. \ref{eq:final_state}). In other cases when $k\neq0$, we can use single RZ gates to correct the additional phases appeared in each spin qubit pair. For example, Bob can apply an ${\rm RZ}(\omega^k)$ gate on his $0$'th qubit to correct the phase and transform $(\ket{00}+\omega^k \ket{11})_{A_0 B_0}$ to $(\ket{00}+ \ket{11})_{A_0 B_0}$. Notably, the correction doesn't require additional communication between Alice and Bob and can be completed fully locally.  Therefore, the X measurement can indeed eliminate the photonic information and project the spin states to multiple EPR states.

\begin{figure}
    \centering
    \includegraphics[width = \linewidth]{ 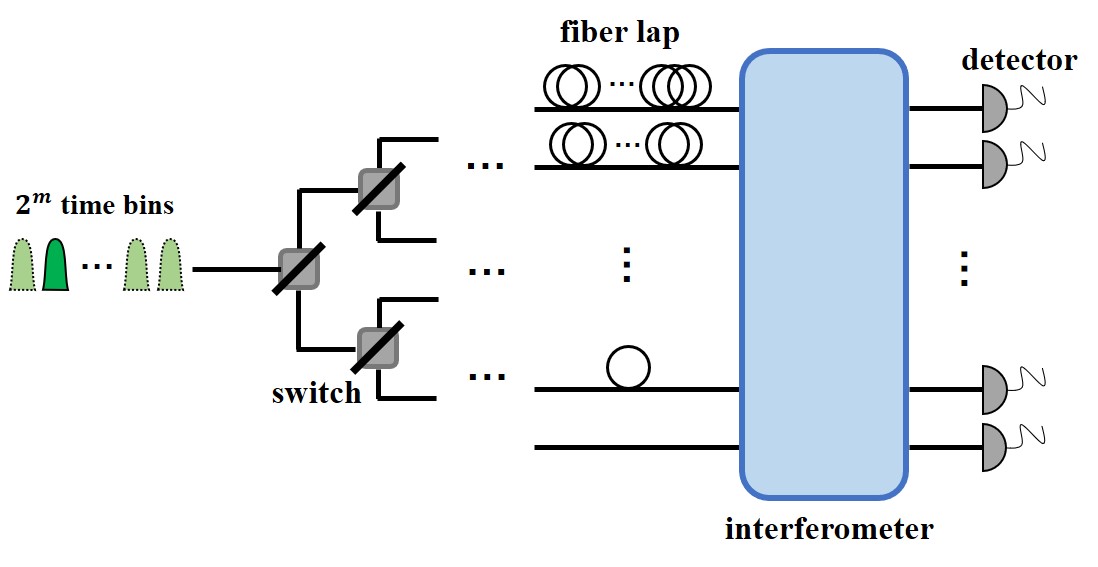}
    \caption{Scheme of the X-basis measurement on the time-bin photonic qudit.}
    \label{fig:X_meas}
\end{figure}

In the X measurement setup (Fig.~\ref{fig:X_meas}), a set of switches are used to distribute the time-bin states to different optical paths. The different paths contain a varying number of fiber loops (depending on the corresponding time-bin state for this path) in order to shift all time-bins to the same temporal mode. In this way, we convert the temporal-mode photonic qudit to a spatial-mode photonic qudit. This allows to use a multimode interferometer to interfere the different spatial states to implement a quantum Fourier transform~\cite{Torma1996,Barak2007,Carolan2015}. Finally, photon detectors will be placed on the output of the interferometer for measurement.

Each single switch will have an effective transmission $\eta_{sw}(1-e_{sw})$. The wrong switching error enters in the transmission since such an error will guide a time-bin to a wrong path, which will fail to align the temporal modes.  Upon detection, this is therefore an heralded error and will be discarded. The photonic qudit will in total pass through $m$ switches and the total transmission will thus be $(\eta_{sw}(1-e_{sw}))^m$.

The fiber loops also result in additional loss due to the longer propagation path of the photon. Importantly, this result in asymmetric loss for the different time-bins since they are experiencing different delays. This would lead to an error in the entanglement generation but it can be completely mitigated by ensuring a matching asymmetry in the initial qudit superposition in the photon generation step. In the simulation, we assume that this is employed to mitigate this error up the level of the amplitude fluctuations that we include (see Eq.~(\ref{eq:quditgen})). The loss experienced in each fiber loop is assumed to be $\eta_{lag}=1\%$ loss.

We model the effect of phase errors in the interferometer as an overall dephasing channel, which causes a decay factor $e^{-\sigma^2_X/2}$ in the non-diagonal elements of the density matrix. In our simulation, we assumed $\sigma_X = 0.1m$ such that this error increases with the number of entangled pairs $m$ since we anticipate that the level of achievable phase stability will decrease with the size of the interferometer.

\section{Decoherence}

We model the decoherence of the spin qubits as a combination of a pure dephasing and generalized amplitude damping channel. 
For qubits with dephasing time $T_p$, the Kraus operators describing a dephasing channel are
\begin{equation}
    \begin{cases}
    &    A_0 = \sqrt{\frac{1+e^{-t/T_p}}{2}}\begin{bmatrix}1& 0\\ 0 & 1 \end{bmatrix} \\
    &    A_1 = \sqrt{\frac{1-e^{-t/T_p}}{2}}\begin{bmatrix}1& 0\\ 0 & -1 \end{bmatrix} .
    \end{cases}
\end{equation}
The Kraus operators for the generalized amplitude damping channel are
\begin{equation}
    \begin{cases}
    &    E_0 = \sqrt{a_\beta}\begin{bmatrix} 1 & 0\\ 0 & e^{-t/2T_1}  \end{bmatrix}\\
    &        E_1 = \sqrt{a_\beta}\begin{bmatrix} 0 & \sqrt{1-e^{-t/T_1}}\\ 0 & 0  \end{bmatrix} \\
    &       E_2 = \sqrt{1-a_\beta}\begin{bmatrix} e^{-t/2T_1}  & 0\\ 0 & 1 \end{bmatrix}\\
    &        E_3 = \sqrt{1-a_\beta}\begin{bmatrix} 0 & 0\\ \sqrt{1-e^{-t/T_1}} & 0  \end{bmatrix},
    \end{cases}
\end{equation}
which will lead to a steady state of the generalize amplitude damping channel
\begin{equation}
    \rho_{\beta} = \begin{bmatrix} a_\beta & 0\\ 0 & 1-a_\beta \end{bmatrix}.
\end{equation}

The decoherence of the spin qubits is thus modelled as a channel
\begin{equation}
    \bar{\rho} = \sum_{i={0,1}, j={0,1,2,3}} (A_i E_j)\rho (A_i E_j)^\dagger
\end{equation}
acting on each of the individual spin qubits. 
In the simulation, we took $T_1=10{\rm ms}$ and $T_p = 5{\rm ms}$. We take $t$ to be the waiting time between the initialization of an atomic qubit and when Alice receives the successful message from Bob after the heralded X-basis measurement. We assume that the time of local operations such as the generation of the qudit and entangling operations is negligible compared to the signalling time between Alice and Bob. Assuming that a time-bin has duration $\approx$ 0.1 $\mu$s, this is justified for distances $L\gtrsim 3$ km. Consequently, When all the entangled pairs are generated simultaneously, Alice's qubits experience a minimum waiting time of $t=2t_0=2L/c$, and Bob's qubits experience a waiting time of $t=L/c$ for the duration of the sucessful message transmitted to Alice. For the all-keep qubit protocol, waiting time for each atomic qubit is determined by Monte Carlo simulation.


\begin{thebibliography}{48}%
    \makeatletter
    \providecommand \@ifxundefined [1]{%
     \@ifx{#1\undefined}
    }%
    \providecommand \@ifnum [1]{%
     \ifnum #1\expandafter \@firstoftwo
     \else \expandafter \@secondoftwo
     \fi
    }%
    \providecommand \@ifx [1]{%
     \ifx #1\expandafter \@firstoftwo
     \else \expandafter \@secondoftwo
     \fi
    }%
    \providecommand \natexlab [1]{#1}%
    \providecommand \enquote  [1]{``#1''}%
    \providecommand \bibnamefont  [1]{#1}%
    \providecommand \bibfnamefont [1]{#1}%
    \providecommand \citenamefont [1]{#1}%
    \providecommand \href@noop [0]{\@secondoftwo}%
    \providecommand \href [0]{\begingroup \@sanitize@url \@href}%
    \providecommand \@href[1]{\@@startlink{#1}\@@href}%
    \providecommand \@@href[1]{\endgroup#1\@@endlink}%
    \providecommand \@sanitize@url [0]{\catcode `\\12\catcode `\$12\catcode
      `\&12\catcode `\#12\catcode `\^12\catcode `\_12\catcode `\%12\relax}%
    \providecommand \@@startlink[1]{}%
    \providecommand \@@endlink[0]{}%
    \providecommand \url  [0]{\begingroup\@sanitize@url \@url }%
    \providecommand \@url [1]{\endgroup\@href {#1}{\urlprefix }}%
    \providecommand \urlprefix  [0]{URL }%
    \providecommand \Eprint [0]{\href }%
    \providecommand \doibase [0]{https://doi.org/}%
    \providecommand \selectlanguage [0]{\@gobble}%
    \providecommand \bibinfo  [0]{\@secondoftwo}%
    \providecommand \bibfield  [0]{\@secondoftwo}%
    \providecommand \translation [1]{[#1]}%
    \providecommand \BibitemOpen [0]{}%
    \providecommand \bibitemStop [0]{}%
    \providecommand \bibitemNoStop [0]{.\EOS\space}%
    \providecommand \EOS [0]{\spacefactor3000\relax}%
    \providecommand \BibitemShut  [1]{\csname bibitem#1\endcsname}%
    \let\auto@bib@innerbib\@empty
    \bibitem [{\citenamefont {Lo}\ \emph {et~al.}(2014)\citenamefont {Lo},
      \citenamefont {Curty},\ and\ \citenamefont {Tamaki}}]{Lo2014}%
      \BibitemOpen
      \bibfield  {author} {\bibinfo {author} {\bibfnamefont {H.-K.}\ \bibnamefont
      {Lo}}, \bibinfo {author} {\bibfnamefont {M.}~\bibnamefont {Curty}},\ and\
      \bibinfo {author} {\bibfnamefont {K.}~\bibnamefont {Tamaki}},\ }\bibfield
      {title} {\bibinfo {title} {Secure quantum key distribution},\ }\href
      {https://doi.org/10.1038/nphoton.2014.149} {\bibfield  {journal} {\bibinfo
      {journal} {Nature Photonics}\ }\textbf {\bibinfo {volume} {8}},\ \bibinfo
      {pages} {595} (\bibinfo {year} {2014})}\BibitemShut {NoStop}%
    \bibitem [{\citenamefont {Pirandola}\ \emph {et~al.}(2020)\citenamefont
      {Pirandola}, \citenamefont {Andersen}, \citenamefont {Banchi}, \citenamefont
      {Berta}, \citenamefont {Bunandar}, \citenamefont {Colbeck}, \citenamefont
      {Englund}, \citenamefont {Gehring}, \citenamefont {Lupo}, \citenamefont
      {Ottaviani}, \citenamefont {Pereira}, \citenamefont {Razavi}, \citenamefont
      {Shaari}, \citenamefont {Tomamichel}, \citenamefont {Usenko}, \citenamefont
      {Vallone}, \citenamefont {Villoresi},\ and\ \citenamefont
      {Wallden}}]{Pirandola2020}%
      \BibitemOpen
      \bibfield  {author} {\bibinfo {author} {\bibfnamefont {S.}~\bibnamefont
      {Pirandola}}, \bibinfo {author} {\bibfnamefont {U.~L.}\ \bibnamefont
      {Andersen}}, \bibinfo {author} {\bibfnamefont {L.}~\bibnamefont {Banchi}},
      \bibinfo {author} {\bibfnamefont {M.}~\bibnamefont {Berta}}, \bibinfo
      {author} {\bibfnamefont {D.}~\bibnamefont {Bunandar}}, \bibinfo {author}
      {\bibfnamefont {R.}~\bibnamefont {Colbeck}}, \bibinfo {author} {\bibfnamefont
      {D.}~\bibnamefont {Englund}}, \bibinfo {author} {\bibfnamefont
      {T.}~\bibnamefont {Gehring}}, \bibinfo {author} {\bibfnamefont
      {C.}~\bibnamefont {Lupo}}, \bibinfo {author} {\bibfnamefont {C.}~\bibnamefont
      {Ottaviani}}, \bibinfo {author} {\bibfnamefont {J.~L.}\ \bibnamefont
      {Pereira}}, \bibinfo {author} {\bibfnamefont {M.}~\bibnamefont {Razavi}},
      \bibinfo {author} {\bibfnamefont {J.~S.}\ \bibnamefont {Shaari}}, \bibinfo
      {author} {\bibfnamefont {M.}~\bibnamefont {Tomamichel}}, \bibinfo {author}
      {\bibfnamefont {V.~C.}\ \bibnamefont {Usenko}}, \bibinfo {author}
      {\bibfnamefont {G.}~\bibnamefont {Vallone}}, \bibinfo {author} {\bibfnamefont
      {P.}~\bibnamefont {Villoresi}},\ and\ \bibinfo {author} {\bibfnamefont
      {P.}~\bibnamefont {Wallden}},\ }\bibfield  {title} {\bibinfo {title}
      {Advances in quantum cryptography},\ }\href
      {https://doi.org/10.1364/AOP.361502} {\bibfield  {journal} {\bibinfo
      {journal} {Adv. Opt. Photon.}\ }\textbf {\bibinfo {volume} {12}},\ \bibinfo
      {pages} {1012} (\bibinfo {year} {2020})}\BibitemShut {NoStop}%
    \bibitem [{\citenamefont {K{\'o}m{\'a}r}\ \emph {et~al.}(2014)\citenamefont
      {K{\'o}m{\'a}r}, \citenamefont {Kessler}, \citenamefont {Bishof},
      \citenamefont {Jiang}, \citenamefont {S{\o}rensen}, \citenamefont {Ye},\ and\
      \citenamefont {Lukin}}]{Komar2014}%
      \BibitemOpen
      \bibfield  {author} {\bibinfo {author} {\bibfnamefont {P.}~\bibnamefont
      {K{\'o}m{\'a}r}}, \bibinfo {author} {\bibfnamefont {E.~M.}\ \bibnamefont
      {Kessler}}, \bibinfo {author} {\bibfnamefont {M.}~\bibnamefont {Bishof}},
      \bibinfo {author} {\bibfnamefont {L.}~\bibnamefont {Jiang}}, \bibinfo
      {author} {\bibfnamefont {A.~S.}\ \bibnamefont {S{\o}rensen}}, \bibinfo
      {author} {\bibfnamefont {J.}~\bibnamefont {Ye}},\ and\ \bibinfo {author}
      {\bibfnamefont {M.~D.}\ \bibnamefont {Lukin}},\ }\bibfield  {title} {\bibinfo
      {title} {A quantum network of clocks},\ }\href
      {https://doi.org/10.1038/nphys3000} {\bibfield  {journal} {\bibinfo
      {journal} {Nature Physics}\ }\textbf {\bibinfo {volume} {10}},\ \bibinfo
      {pages} {582} (\bibinfo {year} {2014})}\BibitemShut {NoStop}%
    \bibitem [{\citenamefont {Khabiboulline}\ \emph
      {et~al.}(2019{\natexlab{a}})\citenamefont {Khabiboulline}, \citenamefont
      {Borregaard}, \citenamefont {De~Greve},\ and\ \citenamefont
      {Lukin}}]{Khabiboulline2019}%
      \BibitemOpen
      \bibfield  {author} {\bibinfo {author} {\bibfnamefont {E.~T.}\ \bibnamefont
      {Khabiboulline}}, \bibinfo {author} {\bibfnamefont {J.}~\bibnamefont
      {Borregaard}}, \bibinfo {author} {\bibfnamefont {K.}~\bibnamefont
      {De~Greve}},\ and\ \bibinfo {author} {\bibfnamefont {M.~D.}\ \bibnamefont
      {Lukin}},\ }\bibfield  {title} {\bibinfo {title} {Optical interferometry with
      quantum networks},\ }\href {https://doi.org/10.1103/PhysRevLett.123.070504}
      {\bibfield  {journal} {\bibinfo  {journal} {Phys. Rev. Lett.}\ }\textbf
      {\bibinfo {volume} {123}},\ \bibinfo {pages} {070504} (\bibinfo {year}
      {2019}{\natexlab{a}})}\BibitemShut {NoStop}%
    \bibitem [{\citenamefont {Guo}\ \emph {et~al.}(2020)\citenamefont {Guo},
      \citenamefont {Breum}, \citenamefont {Borregaard}, \citenamefont {Izumi},
      \citenamefont {Larsen}, \citenamefont {Gehring}, \citenamefont {Christandl},
      \citenamefont {Neergaard-Nielsen},\ and\ \citenamefont {Andersen}}]{Guo2020}%
      \BibitemOpen
      \bibfield  {author} {\bibinfo {author} {\bibfnamefont {X.}~\bibnamefont
      {Guo}}, \bibinfo {author} {\bibfnamefont {C.~R.}\ \bibnamefont {Breum}},
      \bibinfo {author} {\bibfnamefont {J.}~\bibnamefont {Borregaard}}, \bibinfo
      {author} {\bibfnamefont {S.}~\bibnamefont {Izumi}}, \bibinfo {author}
      {\bibfnamefont {M.~V.}\ \bibnamefont {Larsen}}, \bibinfo {author}
      {\bibfnamefont {T.}~\bibnamefont {Gehring}}, \bibinfo {author} {\bibfnamefont
      {M.}~\bibnamefont {Christandl}}, \bibinfo {author} {\bibfnamefont {J.~S.}\
      \bibnamefont {Neergaard-Nielsen}},\ and\ \bibinfo {author} {\bibfnamefont
      {U.~L.}\ \bibnamefont {Andersen}},\ }\bibfield  {title} {\bibinfo {title}
      {Distributed quantum sensing in a continuous-variable entangled network},\
      }\href {https://doi.org/10.1038/s41567-019-0743-x} {\bibfield  {journal}
      {\bibinfo  {journal} {Nature Physics}\ }\textbf {\bibinfo {volume} {16}},\
      \bibinfo {pages} {281} (\bibinfo {year} {2020})}\BibitemShut {NoStop}%
    \bibitem [{\citenamefont {Van~Meter}\ and\ \citenamefont
      {Devitt}(2016)}]{Meter2016}%
      \BibitemOpen
      \bibfield  {author} {\bibinfo {author} {\bibfnamefont {R.}~\bibnamefont
      {Van~Meter}}\ and\ \bibinfo {author} {\bibfnamefont {S.~J.}\ \bibnamefont
      {Devitt}},\ }\bibfield  {title} {\bibinfo {title} {The path to scalable
      distributed quantum computing},\ }\href {https://doi.org/10.1109/MC.2016.291}
      {\bibfield  {journal} {\bibinfo  {journal} {Computer}\ }\textbf {\bibinfo
      {volume} {49}},\ \bibinfo {pages} {31} (\bibinfo {year} {2016})}\BibitemShut
      {NoStop}%
    \bibitem [{\citenamefont {Wehner}\ \emph {et~al.}(2018)\citenamefont {Wehner},
      \citenamefont {Elkouss},\ and\ \citenamefont {Hanson}}]{wehner2018}%
      \BibitemOpen
      \bibfield  {author} {\bibinfo {author} {\bibfnamefont {S.}~\bibnamefont
      {Wehner}}, \bibinfo {author} {\bibfnamefont {D.}~\bibnamefont {Elkouss}},\
      and\ \bibinfo {author} {\bibfnamefont {R.}~\bibnamefont {Hanson}},\
      }\bibfield  {title} {\bibinfo {title} {Quantum internet: A vision for the
      road ahead},\ }\href {https://doi.org/10.1126/science.aam9288} {\bibfield
      {journal} {\bibinfo  {journal} {Science}\ }\textbf {\bibinfo {volume}
      {362}},\ \bibinfo {pages} {eaam9288} (\bibinfo {year} {2018})},\ \Eprint
      {https://arxiv.org/abs/https://www.science.org/doi/pdf/10.1126/science.aam9288}
      {https://www.science.org/doi/pdf/10.1126/science.aam9288} \BibitemShut
      {NoStop}%
    \bibitem [{\citenamefont {Pirandola}\ \emph {et~al.}(2015)\citenamefont
      {Pirandola}, \citenamefont {Eisert}, \citenamefont {Weedbrook}, \citenamefont
      {Furusawa},\ and\ \citenamefont {Braunstein}}]{Pirandola2015}%
      \BibitemOpen
      \bibfield  {author} {\bibinfo {author} {\bibfnamefont {S.}~\bibnamefont
      {Pirandola}}, \bibinfo {author} {\bibfnamefont {J.}~\bibnamefont {Eisert}},
      \bibinfo {author} {\bibfnamefont {C.}~\bibnamefont {Weedbrook}}, \bibinfo
      {author} {\bibfnamefont {A.}~\bibnamefont {Furusawa}},\ and\ \bibinfo
      {author} {\bibfnamefont {S.~L.}\ \bibnamefont {Braunstein}},\ }\bibfield
      {title} {\bibinfo {title} {Advances in quantum teleportation},\ }\href
      {https://doi.org/10.1038/nphoton.2015.154} {\bibfield  {journal} {\bibinfo
      {journal} {Nature Photonics}\ }\textbf {\bibinfo {volume} {9}},\ \bibinfo
      {pages} {641} (\bibinfo {year} {2015})}\BibitemShut {NoStop}%
    \bibitem [{\citenamefont {Simon}\ and\ \citenamefont
      {Irvine}(2003)}]{Simon2003}%
      \BibitemOpen
      \bibfield  {author} {\bibinfo {author} {\bibfnamefont {C.}~\bibnamefont
      {Simon}}\ and\ \bibinfo {author} {\bibfnamefont {W.~T.~M.}\ \bibnamefont
      {Irvine}},\ }\bibfield  {title} {\bibinfo {title} {Robust long-distance
      entanglement and a loophole-free bell test with ions and photons},\ }\href
      {https://doi.org/10.1103/PhysRevLett.91.110405} {\bibfield  {journal}
      {\bibinfo  {journal} {Phys. Rev. Lett.}\ }\textbf {\bibinfo {volume} {91}},\
      \bibinfo {pages} {110405} (\bibinfo {year} {2003})}\BibitemShut {NoStop}%
    \bibitem [{\citenamefont {Hensen}\ \emph {et~al.}(2015)\citenamefont {Hensen},
      \citenamefont {Bernien}, \citenamefont {Dr{\'e}au}, \citenamefont {Reiserer},
      \citenamefont {Kalb}, \citenamefont {Blok}, \citenamefont {Ruitenberg},
      \citenamefont {Vermeulen}, \citenamefont {Schouten}, \citenamefont
      {Abell{\'a}n}, \citenamefont {Amaya}, \citenamefont {Pruneri}, \citenamefont
      {Mitchell}, \citenamefont {Markham}, \citenamefont {Twitchen}, \citenamefont
      {Elkouss}, \citenamefont {Wehner}, \citenamefont {Taminiau},\ and\
      \citenamefont {Hanson}}]{Hensen2015}%
      \BibitemOpen
      \bibfield  {author} {\bibinfo {author} {\bibfnamefont {B.}~\bibnamefont
      {Hensen}}, \bibinfo {author} {\bibfnamefont {H.}~\bibnamefont {Bernien}},
      \bibinfo {author} {\bibfnamefont {A.~E.}\ \bibnamefont {Dr{\'e}au}}, \bibinfo
      {author} {\bibfnamefont {A.}~\bibnamefont {Reiserer}}, \bibinfo {author}
      {\bibfnamefont {N.}~\bibnamefont {Kalb}}, \bibinfo {author} {\bibfnamefont
      {M.~S.}\ \bibnamefont {Blok}}, \bibinfo {author} {\bibfnamefont
      {J.}~\bibnamefont {Ruitenberg}}, \bibinfo {author} {\bibfnamefont {R.~F.~L.}\
      \bibnamefont {Vermeulen}}, \bibinfo {author} {\bibfnamefont {R.~N.}\
      \bibnamefont {Schouten}}, \bibinfo {author} {\bibfnamefont {C.}~\bibnamefont
      {Abell{\'a}n}}, \bibinfo {author} {\bibfnamefont {W.}~\bibnamefont {Amaya}},
      \bibinfo {author} {\bibfnamefont {V.}~\bibnamefont {Pruneri}}, \bibinfo
      {author} {\bibfnamefont {M.~W.}\ \bibnamefont {Mitchell}}, \bibinfo {author}
      {\bibfnamefont {M.}~\bibnamefont {Markham}}, \bibinfo {author} {\bibfnamefont
      {D.~J.}\ \bibnamefont {Twitchen}}, \bibinfo {author} {\bibfnamefont
      {D.}~\bibnamefont {Elkouss}}, \bibinfo {author} {\bibfnamefont
      {S.}~\bibnamefont {Wehner}}, \bibinfo {author} {\bibfnamefont {T.~H.}\
      \bibnamefont {Taminiau}},\ and\ \bibinfo {author} {\bibfnamefont
      {R.}~\bibnamefont {Hanson}},\ }\bibfield  {title} {\bibinfo {title}
      {Loophole-free bell inequality violation using electron spins separated by
      1.3 kilometres},\ }\href {https://doi.org/10.1038/nature15759} {\bibfield
      {journal} {\bibinfo  {journal} {Nature}\ }\textbf {\bibinfo {volume} {526}},\
      \bibinfo {pages} {682} (\bibinfo {year} {2015})}\BibitemShut {NoStop}%
    \bibitem [{\citenamefont {Pompili}\ \emph {et~al.}(2021)\citenamefont
      {Pompili}, \citenamefont {Hermans}, \citenamefont {Baier}, \citenamefont
      {Beukers}, \citenamefont {Humphreys}, \citenamefont {Schouten}, \citenamefont
      {Vermeulen}, \citenamefont {Tiggelman}, \citenamefont {dos Santos~Martins},
      \citenamefont {Dirkse}, \citenamefont {Wehner},\ and\ \citenamefont
      {Hanson}}]{Pompili2021}%
      \BibitemOpen
      \bibfield  {author} {\bibinfo {author} {\bibfnamefont {M.}~\bibnamefont
      {Pompili}}, \bibinfo {author} {\bibfnamefont {S.~L.~N.}\ \bibnamefont
      {Hermans}}, \bibinfo {author} {\bibfnamefont {S.}~\bibnamefont {Baier}},
      \bibinfo {author} {\bibfnamefont {H.~K.~C.}\ \bibnamefont {Beukers}},
      \bibinfo {author} {\bibfnamefont {P.~C.}\ \bibnamefont {Humphreys}}, \bibinfo
      {author} {\bibfnamefont {R.~N.}\ \bibnamefont {Schouten}}, \bibinfo {author}
      {\bibfnamefont {R.~F.~L.}\ \bibnamefont {Vermeulen}}, \bibinfo {author}
      {\bibfnamefont {M.~J.}\ \bibnamefont {Tiggelman}}, \bibinfo {author}
      {\bibfnamefont {L.}~\bibnamefont {dos Santos~Martins}}, \bibinfo {author}
      {\bibfnamefont {B.}~\bibnamefont {Dirkse}}, \bibinfo {author} {\bibfnamefont
      {S.}~\bibnamefont {Wehner}},\ and\ \bibinfo {author} {\bibfnamefont
      {R.}~\bibnamefont {Hanson}},\ }\bibfield  {title} {\bibinfo {title}
      {Realization of a multinode quantum network of remote solid-state qubits},\
      }\href {https://doi.org/10.1126/science.abg1919} {\bibfield  {journal}
      {\bibinfo  {journal} {Science}\ }\textbf {\bibinfo {volume} {372}},\ \bibinfo
      {pages} {259} (\bibinfo {year} {2021})},\ \Eprint
      {https://arxiv.org/abs/https://www.science.org/doi/pdf/10.1126/science.abg1919}
      {https://www.science.org/doi/pdf/10.1126/science.abg1919} \BibitemShut
      {NoStop}%
    \bibitem [{\citenamefont {Langenfeld}\ \emph {et~al.}(2021)\citenamefont
      {Langenfeld}, \citenamefont {Welte}, \citenamefont {Hartung}, \citenamefont
      {Daiss}, \citenamefont {Thomas}, \citenamefont {Morin}, \citenamefont
      {Distante},\ and\ \citenamefont {Rempe}}]{Langenfeld2021}%
      \BibitemOpen
      \bibfield  {author} {\bibinfo {author} {\bibfnamefont {S.}~\bibnamefont
      {Langenfeld}}, \bibinfo {author} {\bibfnamefont {S.}~\bibnamefont {Welte}},
      \bibinfo {author} {\bibfnamefont {L.}~\bibnamefont {Hartung}}, \bibinfo
      {author} {\bibfnamefont {S.}~\bibnamefont {Daiss}}, \bibinfo {author}
      {\bibfnamefont {P.}~\bibnamefont {Thomas}}, \bibinfo {author} {\bibfnamefont
      {O.}~\bibnamefont {Morin}}, \bibinfo {author} {\bibfnamefont
      {E.}~\bibnamefont {Distante}},\ and\ \bibinfo {author} {\bibfnamefont
      {G.}~\bibnamefont {Rempe}},\ }\bibfield  {title} {\bibinfo {title} {Quantum
      teleportation between remote qubit memories with only a single photon as a
      resource},\ }\href {https://doi.org/10.1103/PhysRevLett.126.130502}
      {\bibfield  {journal} {\bibinfo  {journal} {Phys. Rev. Lett.}\ }\textbf
      {\bibinfo {volume} {126}},\ \bibinfo {pages} {130502} (\bibinfo {year}
      {2021})}\BibitemShut {NoStop}%
    \bibitem [{\citenamefont {Lago-Rivera}\ \emph {et~al.}(2021)\citenamefont
      {Lago-Rivera}, \citenamefont {Grandi}, \citenamefont {Rakonjac},
      \citenamefont {Seri},\ and\ \citenamefont {de~Riedmatten}}]{Lago-Rivera2021}%
      \BibitemOpen
      \bibfield  {author} {\bibinfo {author} {\bibfnamefont {D.}~\bibnamefont
      {Lago-Rivera}}, \bibinfo {author} {\bibfnamefont {S.}~\bibnamefont {Grandi}},
      \bibinfo {author} {\bibfnamefont {J.~V.}\ \bibnamefont {Rakonjac}}, \bibinfo
      {author} {\bibfnamefont {A.}~\bibnamefont {Seri}},\ and\ \bibinfo {author}
      {\bibfnamefont {H.}~\bibnamefont {de~Riedmatten}},\ }\bibfield  {title}
      {\bibinfo {title} {Telecom-heralded entanglement between multimode
      solid-state quantum memories},\ }\href
      {https://doi.org/10.1038/s41586-021-03481-8} {\bibfield  {journal} {\bibinfo
      {journal} {Nature}\ }\textbf {\bibinfo {volume} {594}},\ \bibinfo {pages}
      {37} (\bibinfo {year} {2021})}\BibitemShut {NoStop}%
    \bibitem [{\citenamefont {Delteil}\ \emph {et~al.}(2016)\citenamefont
      {Delteil}, \citenamefont {Sun}, \citenamefont {Gao}, \citenamefont {Togan},
      \citenamefont {Faelt},\ and\ \citenamefont {Imamo{\u g}lu}}]{Delteil2016}%
      \BibitemOpen
      \bibfield  {author} {\bibinfo {author} {\bibfnamefont {A.}~\bibnamefont
      {Delteil}}, \bibinfo {author} {\bibfnamefont {Z.}~\bibnamefont {Sun}},
      \bibinfo {author} {\bibfnamefont {W.-b.}\ \bibnamefont {Gao}}, \bibinfo
      {author} {\bibfnamefont {E.}~\bibnamefont {Togan}}, \bibinfo {author}
      {\bibfnamefont {S.}~\bibnamefont {Faelt}},\ and\ \bibinfo {author}
      {\bibfnamefont {A.}~\bibnamefont {Imamo{\u g}lu}},\ }\bibfield  {title}
      {\bibinfo {title} {Generation of heralded entanglement between distant hole
      spins},\ }\href {https://doi.org/10.1038/nphys3605} {\bibfield  {journal}
      {\bibinfo  {journal} {Nature Physics}\ }\textbf {\bibinfo {volume} {12}},\
      \bibinfo {pages} {218} (\bibinfo {year} {2016})}\BibitemShut {NoStop}%
    \bibitem [{\citenamefont {Bennett}\ \emph {et~al.}(1996)\citenamefont
      {Bennett}, \citenamefont {Brassard}, \citenamefont {Popescu}, \citenamefont
      {Schumacher}, \citenamefont {Smolin},\ and\ \citenamefont
      {Wootters}}]{Bennett1996}%
      \BibitemOpen
      \bibfield  {author} {\bibinfo {author} {\bibfnamefont {C.~H.}\ \bibnamefont
      {Bennett}}, \bibinfo {author} {\bibfnamefont {G.}~\bibnamefont {Brassard}},
      \bibinfo {author} {\bibfnamefont {S.}~\bibnamefont {Popescu}}, \bibinfo
      {author} {\bibfnamefont {B.}~\bibnamefont {Schumacher}}, \bibinfo {author}
      {\bibfnamefont {J.~A.}\ \bibnamefont {Smolin}},\ and\ \bibinfo {author}
      {\bibfnamefont {W.~K.}\ \bibnamefont {Wootters}},\ }\bibfield  {title}
      {\bibinfo {title} {Purification of noisy entanglement and faithful
      teleportation via noisy channels},\ }\href
      {https://doi.org/10.1103/PhysRevLett.76.722} {\bibfield  {journal} {\bibinfo
      {journal} {Phys. Rev. Lett.}\ }\textbf {\bibinfo {volume} {76}},\ \bibinfo
      {pages} {722} (\bibinfo {year} {1996})}\BibitemShut {NoStop}%
    \bibitem [{\citenamefont {Deutsch}\ \emph {et~al.}(1996)\citenamefont
      {Deutsch}, \citenamefont {Ekert}, \citenamefont {Jozsa}, \citenamefont
      {Macchiavello}, \citenamefont {Popescu},\ and\ \citenamefont
      {Sanpera}}]{Deutsch1996}%
      \BibitemOpen
      \bibfield  {author} {\bibinfo {author} {\bibfnamefont {D.}~\bibnamefont
      {Deutsch}}, \bibinfo {author} {\bibfnamefont {A.}~\bibnamefont {Ekert}},
      \bibinfo {author} {\bibfnamefont {R.}~\bibnamefont {Jozsa}}, \bibinfo
      {author} {\bibfnamefont {C.}~\bibnamefont {Macchiavello}}, \bibinfo {author}
      {\bibfnamefont {S.}~\bibnamefont {Popescu}},\ and\ \bibinfo {author}
      {\bibfnamefont {A.}~\bibnamefont {Sanpera}},\ }\bibfield  {title} {\bibinfo
      {title} {Quantum privacy amplification and the security of quantum
      cryptography over noisy channels},\ }\href
      {https://doi.org/10.1103/PhysRevLett.77.2818} {\bibfield  {journal} {\bibinfo
       {journal} {Phys. Rev. Lett.}\ }\textbf {\bibinfo {volume} {77}},\ \bibinfo
      {pages} {2818} (\bibinfo {year} {1996})}\BibitemShut {NoStop}%
    \bibitem [{\citenamefont {Fowler}\ \emph {et~al.}(2010)\citenamefont {Fowler},
      \citenamefont {Wang}, \citenamefont {Hill}, \citenamefont {Ladd},
      \citenamefont {Van~Meter},\ and\ \citenamefont {Hollenberg}}]{Fowler2010}%
      \BibitemOpen
      \bibfield  {author} {\bibinfo {author} {\bibfnamefont {A.~G.}\ \bibnamefont
      {Fowler}}, \bibinfo {author} {\bibfnamefont {D.~S.}\ \bibnamefont {Wang}},
      \bibinfo {author} {\bibfnamefont {C.~D.}\ \bibnamefont {Hill}}, \bibinfo
      {author} {\bibfnamefont {T.~D.}\ \bibnamefont {Ladd}}, \bibinfo {author}
      {\bibfnamefont {R.}~\bibnamefont {Van~Meter}},\ and\ \bibinfo {author}
      {\bibfnamefont {L.~C.~L.}\ \bibnamefont {Hollenberg}},\ }\bibfield  {title}
      {\bibinfo {title} {Surface code quantum communication},\ }\href
      {https://doi.org/10.1103/PhysRevLett.104.180503} {\bibfield  {journal}
      {\bibinfo  {journal} {Phys. Rev. Lett.}\ }\textbf {\bibinfo {volume} {104}},\
      \bibinfo {pages} {180503} (\bibinfo {year} {2010})}\BibitemShut {NoStop}%
    \bibitem [{\citenamefont {Munro}\ \emph {et~al.}(2015)\citenamefont {Munro},
      \citenamefont {Azuma}, \citenamefont {Tamaki},\ and\ \citenamefont
      {Nemoto}}]{Munro2015}%
      \BibitemOpen
      \bibfield  {author} {\bibinfo {author} {\bibfnamefont {W.~J.}\ \bibnamefont
      {Munro}}, \bibinfo {author} {\bibfnamefont {K.}~\bibnamefont {Azuma}},
      \bibinfo {author} {\bibfnamefont {K.}~\bibnamefont {Tamaki}},\ and\ \bibinfo
      {author} {\bibfnamefont {K.}~\bibnamefont {Nemoto}},\ }\bibfield  {title}
      {\bibinfo {title} {Inside quantum repeaters},\ }\href
      {https://doi.org/10.1109/JSTQE.2015.2392076} {\bibfield  {journal} {\bibinfo
      {journal} {IEEE Journal of Selected Topics in Quantum Electronics}\ }\textbf
      {\bibinfo {volume} {21}},\ \bibinfo {pages} {78} (\bibinfo {year}
      {2015})}\BibitemShut {NoStop}%
    \bibitem [{\citenamefont {Muralidharan}\ \emph {et~al.}(2016)\citenamefont
      {Muralidharan}, \citenamefont {Li}, \citenamefont {Kim}, \citenamefont
      {L{\"u}tkenhaus}, \citenamefont {Lukin},\ and\ \citenamefont
      {Jiang}}]{Muralidharan2016}%
      \BibitemOpen
      \bibfield  {author} {\bibinfo {author} {\bibfnamefont {S.}~\bibnamefont
      {Muralidharan}}, \bibinfo {author} {\bibfnamefont {L.}~\bibnamefont {Li}},
      \bibinfo {author} {\bibfnamefont {J.}~\bibnamefont {Kim}}, \bibinfo {author}
      {\bibfnamefont {N.}~\bibnamefont {L{\"u}tkenhaus}}, \bibinfo {author}
      {\bibfnamefont {M.~D.}\ \bibnamefont {Lukin}},\ and\ \bibinfo {author}
      {\bibfnamefont {L.}~\bibnamefont {Jiang}},\ }\bibfield  {title} {\bibinfo
      {title} {Optimal architectures for long distance quantum communication},\
      }\href {https://doi.org/10.1038/srep20463} {\bibfield  {journal} {\bibinfo
      {journal} {Scientific Reports}\ }\textbf {\bibinfo {volume} {6}},\ \bibinfo
      {pages} {20463} (\bibinfo {year} {2016})}\BibitemShut {NoStop}%
    \bibitem [{\citenamefont {Childress}\ \emph {et~al.}(2005)\citenamefont
      {Childress}, \citenamefont {Taylor}, \citenamefont {S\o{}rensen},\ and\
      \citenamefont {Lukin}}]{Childress2005}%
      \BibitemOpen
      \bibfield  {author} {\bibinfo {author} {\bibfnamefont {L.}~\bibnamefont
      {Childress}}, \bibinfo {author} {\bibfnamefont {J.~M.}\ \bibnamefont
      {Taylor}}, \bibinfo {author} {\bibfnamefont {A.~S.}\ \bibnamefont
      {S\o{}rensen}},\ and\ \bibinfo {author} {\bibfnamefont {M.~D.}\ \bibnamefont
      {Lukin}},\ }\bibfield  {title} {\bibinfo {title} {Fault-tolerant quantum
      repeaters with minimal physical resources and implementations based on
      single-photon emitters},\ }\href {https://doi.org/10.1103/PhysRevA.72.052330}
      {\bibfield  {journal} {\bibinfo  {journal} {Phys. Rev. A}\ }\textbf {\bibinfo
      {volume} {72}},\ \bibinfo {pages} {052330} (\bibinfo {year}
      {2005})}\BibitemShut {NoStop}%
    \bibitem [{\citenamefont {Kalb}\ \emph {et~al.}(2017)\citenamefont {Kalb},
      \citenamefont {Reiserer}, \citenamefont {Humphreys}, \citenamefont
      {Bakermans}, \citenamefont {Kamerling}, \citenamefont {Nickerson},
      \citenamefont {Benjamin}, \citenamefont {Twitchen}, \citenamefont {Markham},\
      and\ \citenamefont {Hanson}}]{Kalb2017}%
      \BibitemOpen
      \bibfield  {author} {\bibinfo {author} {\bibfnamefont {N.}~\bibnamefont
      {Kalb}}, \bibinfo {author} {\bibfnamefont {A.~A.}\ \bibnamefont {Reiserer}},
      \bibinfo {author} {\bibfnamefont {P.~C.}\ \bibnamefont {Humphreys}}, \bibinfo
      {author} {\bibfnamefont {J.~J.~W.}\ \bibnamefont {Bakermans}}, \bibinfo
      {author} {\bibfnamefont {S.~J.}\ \bibnamefont {Kamerling}}, \bibinfo {author}
      {\bibfnamefont {N.~H.}\ \bibnamefont {Nickerson}}, \bibinfo {author}
      {\bibfnamefont {S.~C.}\ \bibnamefont {Benjamin}}, \bibinfo {author}
      {\bibfnamefont {D.~J.}\ \bibnamefont {Twitchen}}, \bibinfo {author}
      {\bibfnamefont {M.}~\bibnamefont {Markham}},\ and\ \bibinfo {author}
      {\bibfnamefont {R.}~\bibnamefont {Hanson}},\ }\bibfield  {title} {\bibinfo
      {title} {Entanglement distillation between solid-state quantum network
      nodes},\ }\href {https://doi.org/10.1126/science.aan0070} {\bibfield
      {journal} {\bibinfo  {journal} {Science}\ }\textbf {\bibinfo {volume}
      {356}},\ \bibinfo {pages} {928} (\bibinfo {year} {2017})},\ \Eprint
      {https://arxiv.org/abs/https://www.science.org/doi/pdf/10.1126/science.aan0070}
      {https://www.science.org/doi/pdf/10.1126/science.aan0070} \BibitemShut
      {NoStop}%
    \bibitem [{\citenamefont {Jiang}\ \emph {et~al.}(2009)\citenamefont {Jiang},
      \citenamefont {Taylor}, \citenamefont {Nemoto}, \citenamefont {Munro},
      \citenamefont {Van~Meter},\ and\ \citenamefont {Lukin}}]{Liang2009}%
      \BibitemOpen
      \bibfield  {author} {\bibinfo {author} {\bibfnamefont {L.}~\bibnamefont
      {Jiang}}, \bibinfo {author} {\bibfnamefont {J.~M.}\ \bibnamefont {Taylor}},
      \bibinfo {author} {\bibfnamefont {K.}~\bibnamefont {Nemoto}}, \bibinfo
      {author} {\bibfnamefont {W.~J.}\ \bibnamefont {Munro}}, \bibinfo {author}
      {\bibfnamefont {R.}~\bibnamefont {Van~Meter}},\ and\ \bibinfo {author}
      {\bibfnamefont {M.~D.}\ \bibnamefont {Lukin}},\ }\bibfield  {title} {\bibinfo
      {title} {Quantum repeater with encoding},\ }\href
      {https://doi.org/10.1103/PhysRevA.79.032325} {\bibfield  {journal} {\bibinfo
      {journal} {Phys. Rev. A}\ }\textbf {\bibinfo {volume} {79}},\ \bibinfo
      {pages} {032325} (\bibinfo {year} {2009})}\BibitemShut {NoStop}%
    \bibitem [{\citenamefont {Kalb}\ \emph {et~al.}(2018)\citenamefont {Kalb},
      \citenamefont {Humphreys}, \citenamefont {Slim},\ and\ \citenamefont
      {Hanson}}]{Kalb2018}%
      \BibitemOpen
      \bibfield  {author} {\bibinfo {author} {\bibfnamefont {N.}~\bibnamefont
      {Kalb}}, \bibinfo {author} {\bibfnamefont {P.~C.}\ \bibnamefont {Humphreys}},
      \bibinfo {author} {\bibfnamefont {J.~J.}\ \bibnamefont {Slim}},\ and\
      \bibinfo {author} {\bibfnamefont {R.}~\bibnamefont {Hanson}},\ }\bibfield
      {title} {\bibinfo {title} {Dephasing mechanisms of diamond-based nuclear-spin
      memories for quantum networks},\ }\href
      {https://doi.org/10.1103/PhysRevA.97.062330} {\bibfield  {journal} {\bibinfo
      {journal} {Phys. Rev. A}\ }\textbf {\bibinfo {volume} {97}},\ \bibinfo
      {pages} {062330} (\bibinfo {year} {2018})}\BibitemShut {NoStop}%
    \bibitem [{\citenamefont {Bacco}\ \emph {et~al.}(2021)\citenamefont {Bacco},
      \citenamefont {Bulmer}, \citenamefont {Erhard}, \citenamefont {Huber},\ and\
      \citenamefont {Paesani}}]{Bacco2021}%
      \BibitemOpen
      \bibfield  {author} {\bibinfo {author} {\bibfnamefont {D.}~\bibnamefont
      {Bacco}}, \bibinfo {author} {\bibfnamefont {J.~F.~F.}\ \bibnamefont
      {Bulmer}}, \bibinfo {author} {\bibfnamefont {M.}~\bibnamefont {Erhard}},
      \bibinfo {author} {\bibfnamefont {M.}~\bibnamefont {Huber}},\ and\ \bibinfo
      {author} {\bibfnamefont {S.}~\bibnamefont {Paesani}},\ }\bibfield  {title}
      {\bibinfo {title} {Proposal for practical multidimensional quantum
      networks},\ }\href {https://doi.org/10.1103/PhysRevA.104.052618} {\bibfield
      {journal} {\bibinfo  {journal} {Phys. Rev. A}\ }\textbf {\bibinfo {volume}
      {104}},\ \bibinfo {pages} {052618} (\bibinfo {year} {2021})}\BibitemShut
      {NoStop}%
    \bibitem [{\citenamefont {Muralidharan}\ \emph {et~al.}(2018)\citenamefont
      {Muralidharan}, \citenamefont {Zou}, \citenamefont {Li},\ and\ \citenamefont
      {Jiang}}]{Muralidharan2018}%
      \BibitemOpen
      \bibfield  {author} {\bibinfo {author} {\bibfnamefont {S.}~\bibnamefont
      {Muralidharan}}, \bibinfo {author} {\bibfnamefont {C.-L.}\ \bibnamefont
      {Zou}}, \bibinfo {author} {\bibfnamefont {L.}~\bibnamefont {Li}},\ and\
      \bibinfo {author} {\bibfnamefont {L.}~\bibnamefont {Jiang}},\ }\bibfield
      {title} {\bibinfo {title} {One-way quantum repeaters with quantum
      reed-solomon codes},\ }\href {https://doi.org/10.1103/PhysRevA.97.052316}
      {\bibfield  {journal} {\bibinfo  {journal} {Phys. Rev. A}\ }\textbf {\bibinfo
      {volume} {97}},\ \bibinfo {pages} {052316} (\bibinfo {year}
      {2018})}\BibitemShut {NoStop}%
    \bibitem [{\citenamefont {Imany}\ \emph {et~al.}(2019)\citenamefont {Imany},
      \citenamefont {Jaramillo-Villegas}, \citenamefont {Alshaykh}, \citenamefont
      {Lukens}, \citenamefont {Odele}, \citenamefont {Moore}, \citenamefont
      {Leaird}, \citenamefont {Qi},\ and\ \citenamefont {Weiner}}]{Imany2019}%
      \BibitemOpen
      \bibfield  {author} {\bibinfo {author} {\bibfnamefont {P.}~\bibnamefont
      {Imany}}, \bibinfo {author} {\bibfnamefont {J.~A.}\ \bibnamefont
      {Jaramillo-Villegas}}, \bibinfo {author} {\bibfnamefont {M.~S.}\ \bibnamefont
      {Alshaykh}}, \bibinfo {author} {\bibfnamefont {J.~M.}\ \bibnamefont
      {Lukens}}, \bibinfo {author} {\bibfnamefont {O.~D.}\ \bibnamefont {Odele}},
      \bibinfo {author} {\bibfnamefont {A.~J.}\ \bibnamefont {Moore}}, \bibinfo
      {author} {\bibfnamefont {D.~E.}\ \bibnamefont {Leaird}}, \bibinfo {author}
      {\bibfnamefont {M.}~\bibnamefont {Qi}},\ and\ \bibinfo {author}
      {\bibfnamefont {A.~M.}\ \bibnamefont {Weiner}},\ }\bibfield  {title}
      {\bibinfo {title} {High-dimensional optical quantum logic in large
      operational spaces},\ }\href {https://doi.org/10.1038/s41534-019-0173-8}
      {\bibfield  {journal} {\bibinfo  {journal} {npj Quantum Information}\
      }\textbf {\bibinfo {volume} {5}},\ \bibinfo {pages} {59} (\bibinfo {year}
      {2019})}\BibitemShut {NoStop}%
    \bibitem [{\citenamefont {Bell}\ \emph {et~al.}(2022)\citenamefont {Bell},
      \citenamefont {Bulmer}, \citenamefont {Jones}, \citenamefont {Paesani},
      \citenamefont {McCutcheon},\ and\ \citenamefont {Laing}}]{Bell2022}%
      \BibitemOpen
      \bibfield  {author} {\bibinfo {author} {\bibfnamefont {T.~J.}\ \bibnamefont
      {Bell}}, \bibinfo {author} {\bibfnamefont {J.~F.~F.}\ \bibnamefont {Bulmer}},
      \bibinfo {author} {\bibfnamefont {A.~E.}\ \bibnamefont {Jones}}, \bibinfo
      {author} {\bibfnamefont {S.}~\bibnamefont {Paesani}}, \bibinfo {author}
      {\bibfnamefont {D.~P.~S.}\ \bibnamefont {McCutcheon}},\ and\ \bibinfo
      {author} {\bibfnamefont {A.}~\bibnamefont {Laing}},\ }\bibfield  {title}
      {\bibinfo {title} {Protocol for generation of high-dimensional entanglement
      from an array of non-interacting photon emitters},\ }\href
      {https://doi.org/10.1088/1367-2630/ac475d} {\bibfield  {journal} {\bibinfo
      {journal} {New Journal of Physics}\ }\textbf {\bibinfo {volume} {24}},\
      \bibinfo {pages} {013032} (\bibinfo {year} {2022})}\BibitemShut {NoStop}%
    \bibitem [{\citenamefont {Duan}\ and\ \citenamefont {Kimble}(2004)}]{Duan2004}%
      \BibitemOpen
      \bibfield  {author} {\bibinfo {author} {\bibfnamefont {L.-M.}\ \bibnamefont
      {Duan}}\ and\ \bibinfo {author} {\bibfnamefont {H.~J.}\ \bibnamefont
      {Kimble}},\ }\bibfield  {title} {\bibinfo {title} {Scalable photonic quantum
      computation through cavity-assisted interactions},\ }\href
      {https://doi.org/10.1103/PhysRevLett.92.127902} {\bibfield  {journal}
      {\bibinfo  {journal} {Phys. Rev. Lett.}\ }\textbf {\bibinfo {volume} {92}},\
      \bibinfo {pages} {127902} (\bibinfo {year} {2004})}\BibitemShut {NoStop}%
    \bibitem [{\citenamefont {Bhaskar}\ \emph {et~al.}(2020)\citenamefont
      {Bhaskar}, \citenamefont {Riedinger}, \citenamefont {Machielse},
      \citenamefont {Levonian}, \citenamefont {Nguyen}, \citenamefont {Knall},
      \citenamefont {Park}, \citenamefont {Englund}, \citenamefont {Lon{\v c}ar},
      \citenamefont {Sukachev},\ and\ \citenamefont {Lukin}}]{Bhaskar2020}%
      \BibitemOpen
      \bibfield  {author} {\bibinfo {author} {\bibfnamefont {M.~K.}\ \bibnamefont
      {Bhaskar}}, \bibinfo {author} {\bibfnamefont {R.}~\bibnamefont {Riedinger}},
      \bibinfo {author} {\bibfnamefont {B.}~\bibnamefont {Machielse}}, \bibinfo
      {author} {\bibfnamefont {D.~S.}\ \bibnamefont {Levonian}}, \bibinfo {author}
      {\bibfnamefont {C.~T.}\ \bibnamefont {Nguyen}}, \bibinfo {author}
      {\bibfnamefont {E.~N.}\ \bibnamefont {Knall}}, \bibinfo {author}
      {\bibfnamefont {H.}~\bibnamefont {Park}}, \bibinfo {author} {\bibfnamefont
      {D.}~\bibnamefont {Englund}}, \bibinfo {author} {\bibfnamefont
      {M.}~\bibnamefont {Lon{\v c}ar}}, \bibinfo {author} {\bibfnamefont {D.~D.}\
      \bibnamefont {Sukachev}},\ and\ \bibinfo {author} {\bibfnamefont {M.~D.}\
      \bibnamefont {Lukin}},\ }\bibfield  {title} {\bibinfo {title} {Experimental
      demonstration of memory-enhanced quantum communication},\ }\href
      {https://doi.org/10.1038/s41586-020-2103-5} {\bibfield  {journal} {\bibinfo
      {journal} {Nature}\ }\textbf {\bibinfo {volume} {580}},\ \bibinfo {pages}
      {60} (\bibinfo {year} {2020})}\BibitemShut {NoStop}%
    \bibitem [{\citenamefont {Bernardes}\ \emph {et~al.}(2011)\citenamefont
      {Bernardes}, \citenamefont {Praxmeyer},\ and\ \citenamefont {van
      Loock}}]{Bernardes2011}%
      \BibitemOpen
      \bibfield  {author} {\bibinfo {author} {\bibfnamefont {N.~K.}\ \bibnamefont
      {Bernardes}}, \bibinfo {author} {\bibfnamefont {L.}~\bibnamefont
      {Praxmeyer}},\ and\ \bibinfo {author} {\bibfnamefont {P.}~\bibnamefont {van
      Loock}},\ }\bibfield  {title} {\bibinfo {title} {Rate analysis for a hybrid
      quantum repeater},\ }\href {https://doi.org/10.1103/PhysRevA.83.012323}
      {\bibfield  {journal} {\bibinfo  {journal} {Phys. Rev. A}\ }\textbf {\bibinfo
      {volume} {83}},\ \bibinfo {pages} {012323} (\bibinfo {year}
      {2011})}\BibitemShut {NoStop}%
    \bibitem [{\citenamefont {Reiserer}\ \emph {et~al.}(2014)\citenamefont
      {Reiserer}, \citenamefont {Kalb}, \citenamefont {Rempe},\ and\ \citenamefont
      {Ritter}}]{Reiserer2014}%
      \BibitemOpen
      \bibfield  {author} {\bibinfo {author} {\bibfnamefont {A.}~\bibnamefont
      {Reiserer}}, \bibinfo {author} {\bibfnamefont {N.}~\bibnamefont {Kalb}},
      \bibinfo {author} {\bibfnamefont {G.}~\bibnamefont {Rempe}},\ and\ \bibinfo
      {author} {\bibfnamefont {S.}~\bibnamefont {Ritter}},\ }\bibfield  {title}
      {\bibinfo {title} {A quantum gate between a flying optical photon and a
      single trapped atom},\ }\href {https://doi.org/10.1038/nature13177}
      {\bibfield  {journal} {\bibinfo  {journal} {Nature}\ }\textbf {\bibinfo
      {volume} {508}},\ \bibinfo {pages} {237} (\bibinfo {year}
      {2014})}\BibitemShut {NoStop}%
    \bibitem [{\citenamefont {Knall}\ \emph {et~al.}()\citenamefont {Knall},
      \citenamefont {Knaut}, \citenamefont {Bekenstein}, \citenamefont {Assumpcao},
      \citenamefont {Stroganov}, \citenamefont {Gong}, \citenamefont {Huan},
      \citenamefont {Stas}, \citenamefont {Machielse}, \citenamefont {Chalupnik},
      \citenamefont {Levonian}, \citenamefont {Suleymanzade}, \citenamefont
      {Riedinger}, \citenamefont {Park}, \citenamefont {Lon{\v c}ar}, \citenamefont
      {Bhaskar},\ and\ \citenamefont {Lukin}}]{Knall2022}%
      \BibitemOpen
      \bibfield  {author} {\bibinfo {author} {\bibfnamefont {E.~N.}\ \bibnamefont
      {Knall}}, \bibinfo {author} {\bibfnamefont {C.~M.}\ \bibnamefont {Knaut}},
      \bibinfo {author} {\bibfnamefont {R.}~\bibnamefont {Bekenstein}}, \bibinfo
      {author} {\bibfnamefont {D.~R.}\ \bibnamefont {Assumpcao}}, \bibinfo {author}
      {\bibfnamefont {P.~L.}\ \bibnamefont {Stroganov}}, \bibinfo {author}
      {\bibfnamefont {W.}~\bibnamefont {Gong}}, \bibinfo {author} {\bibfnamefont
      {Y.~Q.}\ \bibnamefont {Huan}}, \bibinfo {author} {\bibfnamefont {P.-J.}\
      \bibnamefont {Stas}}, \bibinfo {author} {\bibfnamefont {B.}~\bibnamefont
      {Machielse}}, \bibinfo {author} {\bibfnamefont {M.}~\bibnamefont
      {Chalupnik}}, \bibinfo {author} {\bibfnamefont {D.}~\bibnamefont {Levonian}},
      \bibinfo {author} {\bibfnamefont {A.}~\bibnamefont {Suleymanzade}}, \bibinfo
      {author} {\bibfnamefont {R.}~\bibnamefont {Riedinger}}, \bibinfo {author}
      {\bibfnamefont {H.}~\bibnamefont {Park}}, \bibinfo {author} {\bibfnamefont
      {M.}~\bibnamefont {Lon{\v c}ar}}, \bibinfo {author} {\bibfnamefont {M.~K.}\
      \bibnamefont {Bhaskar}},\ and\ \bibinfo {author} {\bibfnamefont {M.~D.}\
      \bibnamefont {Lukin}},\ }\bibfield  {title} {\bibinfo {title} {Efficient
      source of shaped single photons based on an integrated diamond nanophotonic
      system},\ }\href@noop {} {\bibinfo  {journal} {arXiv:2201.02731}\
      }\BibitemShut {NoStop}%
    \bibitem [{\citenamefont {Bradac}\ \emph {et~al.}(2019)\citenamefont {Bradac},
      \citenamefont {Gao}, \citenamefont {Forneris}, \citenamefont {Trusheim},\
      and\ \citenamefont {Aharonovich}}]{Bradac2019}%
      \BibitemOpen
    \bibfield  {journal} {  }\bibfield  {author} {\bibinfo {author} {\bibfnamefont
      {C.}~\bibnamefont {Bradac}}, \bibinfo {author} {\bibfnamefont
      {W.}~\bibnamefont {Gao}}, \bibinfo {author} {\bibfnamefont {J.}~\bibnamefont
      {Forneris}}, \bibinfo {author} {\bibfnamefont {M.~E.}\ \bibnamefont
      {Trusheim}},\ and\ \bibinfo {author} {\bibfnamefont {I.}~\bibnamefont
      {Aharonovich}},\ }\bibfield  {title} {\bibinfo {title} {Quantum nanophotonics
      with group iv defects in diamond},\ }\href
      {https://doi.org/10.1038/s41467-019-13332-w} {\bibfield  {journal} {\bibinfo
      {journal} {Nature Communications}\ }\textbf {\bibinfo {volume} {10}},\
      \bibinfo {pages} {5625} (\bibinfo {year} {2019})}\BibitemShut {NoStop}%
    \bibitem [{\citenamefont {Barak}\ and\ \citenamefont
      {Ben-Aryeh}(2007)}]{Barak2007}%
      \BibitemOpen
      \bibfield  {author} {\bibinfo {author} {\bibfnamefont {R.}~\bibnamefont
      {Barak}}\ and\ \bibinfo {author} {\bibfnamefont {Y.}~\bibnamefont
      {Ben-Aryeh}},\ }\bibfield  {title} {\bibinfo {title} {Quantum fast fourier
      transform and quantum computation by linear optics},\ }\href
      {https://doi.org/10.1364/JOSAB.24.000231} {\bibfield  {journal} {\bibinfo
      {journal} {J. Opt. Soc. Am. B}\ }\textbf {\bibinfo {volume} {24}},\ \bibinfo
      {pages} {231} (\bibinfo {year} {2007})}\BibitemShut {NoStop}%
    \bibitem [{\citenamefont {Khabiboulline}\ \emph
      {et~al.}(2019{\natexlab{b}})\citenamefont {Khabiboulline}, \citenamefont
      {Borregaard}, \citenamefont {De~Greve},\ and\ \citenamefont
      {Lukin}}]{Khabiboulline2019pra}%
      \BibitemOpen
      \bibfield  {author} {\bibinfo {author} {\bibfnamefont {E.~T.}\ \bibnamefont
      {Khabiboulline}}, \bibinfo {author} {\bibfnamefont {J.}~\bibnamefont
      {Borregaard}}, \bibinfo {author} {\bibfnamefont {K.}~\bibnamefont
      {De~Greve}},\ and\ \bibinfo {author} {\bibfnamefont {M.~D.}\ \bibnamefont
      {Lukin}},\ }\bibfield  {title} {\bibinfo {title} {Quantum-assisted telescope
      arrays},\ }\href {https://doi.org/10.1103/PhysRevA.100.022316} {\bibfield
      {journal} {\bibinfo  {journal} {Phys. Rev. A}\ }\textbf {\bibinfo {volume}
      {100}},\ \bibinfo {pages} {022316} (\bibinfo {year}
      {2019}{\natexlab{b}})}\BibitemShut {NoStop}%
    \bibitem [{\citenamefont {Nguyen}\ \emph {et~al.}(2019)\citenamefont {Nguyen},
      \citenamefont {Sukachev}, \citenamefont {Bhaskar}, \citenamefont {Machielse},
      \citenamefont {Levonian}, \citenamefont {Knall}, \citenamefont {Stroganov},
      \citenamefont {Riedinger}, \citenamefont {Park}, \citenamefont {Loncar},\
      and\ \citenamefont {Lukin}}]{Nguyen2019}%
      \BibitemOpen
      \bibfield  {author} {\bibinfo {author} {\bibfnamefont {C.~T.}\ \bibnamefont
      {Nguyen}}, \bibinfo {author} {\bibfnamefont {D.~D.}\ \bibnamefont
      {Sukachev}}, \bibinfo {author} {\bibfnamefont {M.~K.}\ \bibnamefont
      {Bhaskar}}, \bibinfo {author} {\bibfnamefont {B.}~\bibnamefont {Machielse}},
      \bibinfo {author} {\bibfnamefont {D.~S.}\ \bibnamefont {Levonian}}, \bibinfo
      {author} {\bibfnamefont {E.~N.}\ \bibnamefont {Knall}}, \bibinfo {author}
      {\bibfnamefont {P.}~\bibnamefont {Stroganov}}, \bibinfo {author}
      {\bibfnamefont {R.}~\bibnamefont {Riedinger}}, \bibinfo {author}
      {\bibfnamefont {H.}~\bibnamefont {Park}}, \bibinfo {author} {\bibfnamefont
      {M.}~\bibnamefont {Loncar}},\ and\ \bibinfo {author} {\bibfnamefont {M.~D.}\
      \bibnamefont {Lukin}},\ }\bibfield  {title} {\bibinfo {title} {Quantum
      network nodes based on diamond qubits with an efficient nanophotonic
      interface},\ }\href {https://doi.org/10.1103/PhysRevLett.123.183602}
      {\bibfield  {journal} {\bibinfo  {journal} {Phys. Rev. Lett.}\ }\textbf
      {\bibinfo {volume} {123}},\ \bibinfo {pages} {183602} (\bibinfo {year}
      {2019})}\BibitemShut {NoStop}%
    \bibitem [{\citenamefont {Dordevic}\ \emph {et~al.}(2021)\citenamefont
      {Doraevic}, \citenamefont {Samutpraphoot}, \citenamefont {Ocola},
      \citenamefont {Bernien}, \citenamefont {Grinkemeyer}, \citenamefont
      {Dimitrova}, \citenamefont {Vuleti{\'c}},\ and\ \citenamefont
      {Lukin}}]{Tamara2021}%
      \BibitemOpen
      \bibfield  {author} {\bibinfo {author} {\bibfnamefont {T.}~\bibnamefont
      {Doraevic}}, \bibinfo {author} {\bibfnamefont {P.}~\bibnamefont
      {Samutpraphoot}}, \bibinfo {author} {\bibfnamefont {P.~L.}\ \bibnamefont
      {Ocola}}, \bibinfo {author} {\bibfnamefont {H.}~\bibnamefont {Bernien}},
      \bibinfo {author} {\bibfnamefont {B.}~\bibnamefont {Grinkemeyer}}, \bibinfo
      {author} {\bibfnamefont {I.}~\bibnamefont {Dimitrova}}, \bibinfo {author}
      {\bibfnamefont {V.}~\bibnamefont {Vuletic}},\ and\ \bibinfo {author} 
      {\bibfnamefont {M.~D.}\ \bibnamefont {Lukin}},\ }\bibfield  {title} {\bibinfo
      {title} {Entanglement transport and a nanophotonic interface for atoms in
      optical tweezers},\ }\href {https://doi.org/10.1126/science.abi9917}
      {\bibfield  {journal} {\bibinfo  {journal} {Science}\ }\textbf {\bibinfo
      {volume} {373}},\ \bibinfo {pages} {1511} (\bibinfo {year} {2021})},\ \Eprint
      {https://arxiv.org/abs/https://www.science.org/doi/pdf/10.1126/science.abi9917}
      {https://www.science.org/doi/pdf/10.1126/science.abi9917} \BibitemShut
      {NoStop}%
    \bibitem [{\citenamefont {Borregaard}\ \emph {et~al.}(2015)\citenamefont
      {Borregaard}, \citenamefont {K\'om\'ar}, \citenamefont {Kessler},
      \citenamefont {Lukin},\ and\ \citenamefont
      {S\o{}rensen}}]{Borregaard2015pra}%
      \BibitemOpen
      \bibfield  {author} {\bibinfo {author} {\bibfnamefont {J.}~\bibnamefont
      {Borregaard}}, \bibinfo {author} {\bibfnamefont {P.}~\bibnamefont
      {K\'om\'ar}}, \bibinfo {author} {\bibfnamefont {E.~M.}\ \bibnamefont
      {Kessler}}, \bibinfo {author} {\bibfnamefont {M.~D.}\ \bibnamefont {Lukin}},\
      and\ \bibinfo {author} {\bibfnamefont {A.~S.}\ \bibnamefont {S\o{}rensen}},\
      }\bibfield  {title} {\bibinfo {title} {Long-distance entanglement
      distribution using individual atoms in optical cavities},\ }\href
      {https://doi.org/10.1103/PhysRevA.92.012307} {\bibfield  {journal} {\bibinfo
      {journal} {Phys. Rev. A}\ }\textbf {\bibinfo {volume} {92}},\ \bibinfo
      {pages} {012307} (\bibinfo {year} {2015})}\BibitemShut {NoStop}%
    \bibitem [{\citenamefont {Uphoff}\ \emph {et~al.}(2016)\citenamefont {Uphoff},
      \citenamefont {Brekenfeld}, \citenamefont {Rempe},\ and\ \citenamefont
      {Ritter}}]{Uphoff2016}%
      \BibitemOpen
      \bibfield  {author} {\bibinfo {author} {\bibfnamefont {M.}~\bibnamefont
      {Uphoff}}, \bibinfo {author} {\bibfnamefont {M.}~\bibnamefont {Brekenfeld}},
      \bibinfo {author} {\bibfnamefont {G.}~\bibnamefont {Rempe}},\ and\ \bibinfo
      {author} {\bibfnamefont {S.}~\bibnamefont {Ritter}},\ }\bibfield  {title}
      {\bibinfo {title} {An integrated quantum repeater at telecom wavelength with
      single atoms in optical fiber cavities},\ }\href
      {https://doi.org/10.1007/s00340-015-6299-2} {\bibfield  {journal} {\bibinfo
      {journal} {Applied Physics B}\ }\textbf {\bibinfo {volume} {122}},\ \bibinfo
      {pages} {46} (\bibinfo {year} {2016})}\BibitemShut {NoStop}%
    \bibitem [{\citenamefont {Huie}\ \emph {et~al.}(2021)\citenamefont {Huie},
      \citenamefont {Menon}, \citenamefont {Bernien},\ and\ \citenamefont
      {Covey}}]{Huie2021}%
      \BibitemOpen
      \bibfield  {author} {\bibinfo {author} {\bibfnamefont {W.}~\bibnamefont
      {Huie}}, \bibinfo {author} {\bibfnamefont {S.~G.}\ \bibnamefont {Menon}},
      \bibinfo {author} {\bibfnamefont {H.}~\bibnamefont {Bernien}},\ and\ \bibinfo
      {author} {\bibfnamefont {J.~P.}\ \bibnamefont {Covey}},\ }\bibfield  {title}
      {\bibinfo {title} {Multiplexed telecommunication-band quantum networking with
      atom arrays in optical cavities},\ }\href
      {https://doi.org/10.1103/PhysRevResearch.3.043154} {\bibfield  {journal}
      {\bibinfo  {journal} {Phys. Rev. Research}\ }\textbf {\bibinfo {volume}
      {3}},\ \bibinfo {pages} {043154} (\bibinfo {year} {2021})}\BibitemShut
      {NoStop}%
    \bibitem [{\citenamefont {Borregaard}\ \emph {et~al.}(2020)\citenamefont
      {Borregaard}, \citenamefont {Pichler}, \citenamefont {Schr\"oder},
      \citenamefont {Lukin}, \citenamefont {Lodahl},\ and\ \citenamefont
      {S\o{}rensen}}]{borregaard2020}%
      \BibitemOpen
      \bibfield  {author} {\bibinfo {author} {\bibfnamefont {J.}~\bibnamefont
      {Borregaard}}, \bibinfo {author} {\bibfnamefont {H.}~\bibnamefont {Pichler}},
      \bibinfo {author} {\bibfnamefont {T.}~\bibnamefont {Schr\"oder}}, \bibinfo
      {author} {\bibfnamefont {M.~D.}\ \bibnamefont {Lukin}}, \bibinfo {author}
      {\bibfnamefont {P.}~\bibnamefont {Lodahl}},\ and\ \bibinfo {author}
      {\bibfnamefont {A.~S.}\ \bibnamefont {S\o{}rensen}},\ }\bibfield  {title}
      {\bibinfo {title} {One-way quantum repeater based on near-deterministic
      photon-emitter interfaces},\ }\href
      {https://doi.org/10.1103/PhysRevX.10.021071} {\bibfield  {journal} {\bibinfo
      {journal} {Phys. Rev. X}\ }\textbf {\bibinfo {volume} {10}},\ \bibinfo
      {pages} {021071} (\bibinfo {year} {2020})}\BibitemShut {NoStop}%
    \bibitem [{\citenamefont {Collins}\ \emph {et~al.}(2007)\citenamefont
      {Collins}, \citenamefont {Jenkins}, \citenamefont {Kuzmich},\ and\
      \citenamefont {Kennedy}}]{Collins2007}%
      \BibitemOpen
      \bibfield  {author} {\bibinfo {author} {\bibfnamefont {O.~A.}\ \bibnamefont
      {Collins}}, \bibinfo {author} {\bibfnamefont {S.~D.}\ \bibnamefont
      {Jenkins}}, \bibinfo {author} {\bibfnamefont {A.}~\bibnamefont {Kuzmich}},\
      and\ \bibinfo {author} {\bibfnamefont {T.~A.~B.}\ \bibnamefont {Kennedy}},\
      }\bibfield  {title} {\bibinfo {title} {Multiplexed memory-insensitive quantum
      repeaters},\ }\href {https://doi.org/10.1103/PhysRevLett.98.060502}
      {\bibfield  {journal} {\bibinfo  {journal} {Phys. Rev. Lett.}\ }\textbf
      {\bibinfo {volume} {98}},\ \bibinfo {pages} {060502} (\bibinfo {year}
      {2007})}\BibitemShut {NoStop}%
    \bibitem [{\citenamefont {Yin}\ \emph {et~al.}(2017)\citenamefont {Yin},
      \citenamefont {Cao}, \citenamefont {Li}, \citenamefont {Liao}, \citenamefont
      {Zhang}, \citenamefont {Ren}, \citenamefont {Cai}, \citenamefont {Liu},
      \citenamefont {Li}, \citenamefont {Dai}, \citenamefont {Li}, \citenamefont
      {Lu}, \citenamefont {Gong}, \citenamefont {Xu}, \citenamefont {Li},
      \citenamefont {Li}, \citenamefont {Yin}, \citenamefont {Jiang}, \citenamefont
      {Li}, \citenamefont {Jia}, \citenamefont {Ren}, \citenamefont {He},
      \citenamefont {Zhou}, \citenamefont {Zhang}, \citenamefont {Wang},
      \citenamefont {Chang}, \citenamefont {Zhu}, \citenamefont {Liu},
      \citenamefont {Chen}, \citenamefont {Lu}, \citenamefont {Shu}, \citenamefont
      {Peng}, \citenamefont {Wang},\ and\ \citenamefont {Pan}}]{Juan2017}%
      \BibitemOpen
      \bibfield  {author} {\bibinfo {author} {\bibfnamefont {J.}~\bibnamefont
      {Yin}}, \bibinfo {author} {\bibfnamefont {Y.}~\bibnamefont {Cao}}, \bibinfo
      {author} {\bibfnamefont {Y.-H.}\ \bibnamefont {Li}}, \bibinfo {author}
      {\bibfnamefont {S.-K.}\ \bibnamefont {Liao}}, \bibinfo {author}
      {\bibfnamefont {L.}~\bibnamefont {Zhang}}, \bibinfo {author} {\bibfnamefont
      {J.-G.}\ \bibnamefont {Ren}}, \bibinfo {author} {\bibfnamefont {W.-Q.}\
      \bibnamefont {Cai}}, \bibinfo {author} {\bibfnamefont {W.-Y.}\ \bibnamefont
      {Liu}}, \bibinfo {author} {\bibfnamefont {B.}~\bibnamefont {Li}}, \bibinfo
      {author} {\bibfnamefont {H.}~\bibnamefont {Dai}}, \bibinfo {author}
      {\bibfnamefont {G.-B.}\ \bibnamefont {Li}}, \bibinfo {author} {\bibfnamefont
      {Q.-M.}\ \bibnamefont {Lu}}, \bibinfo {author} {\bibfnamefont {Y.-H.}\
      \bibnamefont {Gong}}, \bibinfo {author} {\bibfnamefont {Y.}~\bibnamefont
      {Xu}}, \bibinfo {author} {\bibfnamefont {S.-L.}\ \bibnamefont {Li}}, \bibinfo
      {author} {\bibfnamefont {F.-Z.}\ \bibnamefont {Li}}, \bibinfo {author}
      {\bibfnamefont {Y.-Y.}\ \bibnamefont {Yin}}, \bibinfo {author} {\bibfnamefont
      {Z.-Q.}\ \bibnamefont {Jiang}}, \bibinfo {author} {\bibfnamefont
      {M.}~\bibnamefont {Li}}, \bibinfo {author} {\bibfnamefont {J.-J.}\
      \bibnamefont {Jia}}, \bibinfo {author} {\bibfnamefont {G.}~\bibnamefont
      {Ren}}, \bibinfo {author} {\bibfnamefont {D.}~\bibnamefont {He}}, \bibinfo
      {author} {\bibfnamefont {Y.-L.}\ \bibnamefont {Zhou}}, \bibinfo {author}
      {\bibfnamefont {X.-X.}\ \bibnamefont {Zhang}}, \bibinfo {author}
      {\bibfnamefont {N.}~\bibnamefont {Wang}}, \bibinfo {author} {\bibfnamefont
      {X.}~\bibnamefont {Chang}}, \bibinfo {author} {\bibfnamefont {Z.-C.}\
      \bibnamefont {Zhu}}, \bibinfo {author} {\bibfnamefont {N.-L.}\ \bibnamefont
      {Liu}}, \bibinfo {author} {\bibfnamefont {Y.-A.}\ \bibnamefont {Chen}},
      \bibinfo {author} {\bibfnamefont {C.-Y.}\ \bibnamefont {Lu}}, \bibinfo
      {author} {\bibfnamefont {R.}~\bibnamefont {Shu}}, \bibinfo {author}
      {\bibfnamefont {C.-Z.}\ \bibnamefont {Peng}}, \bibinfo {author}
      {\bibfnamefont {J.-Y.}\ \bibnamefont {Wang}},\ and\ \bibinfo {author}
      {\bibfnamefont {J.-W.}\ \bibnamefont {Pan}},\ }\bibfield  {title} {\bibinfo
      {title} {Satellite-based entanglement distribution over 1200 kilometers},\
      }\href {https://doi.org/10.1126/science.aan3211} {\bibfield  {journal}
      {\bibinfo  {journal} {Science}\ }\textbf {\bibinfo {volume} {356}},\ \bibinfo
      {pages} {1140} (\bibinfo {year} {2017})},\ \Eprint
      {https://arxiv.org/abs/https://www.science.org/doi/pdf/10.1126/science.aan3211}
      {https://www.science.org/doi/pdf/10.1126/science.aan3211} \BibitemShut
      {NoStop}%
    \bibitem [{\citenamefont {Dalibard}\ \emph {et~al.}(1992)\citenamefont
      {Dalibard}, \citenamefont {Castin},\ and\ \citenamefont
      {Molmer}}]{Dalibard1992}%
      \BibitemOpen
      \bibfield  {author} {\bibinfo {author} {\bibfnamefont {J.}~\bibnamefont
      {Dalibard}}, \bibinfo {author} {\bibfnamefont {Y.}~\bibnamefont {Castin}},\
      and\ \bibinfo {author} {\bibfnamefont {K.}~\bibnamefont {Molmer}},\
      }\bibfield  {title} {\bibinfo {title} {Wave-function approach to dissipative
      processes in quantum optics},\ }\href
      {https://doi.org/10.1103/PhysRevLett.68.580} {\bibfield  {journal} {\bibinfo
      {journal} {Phys. Rev. Lett.}\ }\textbf {\bibinfo {volume} {68}},\ \bibinfo
      {pages} {580} (\bibinfo {year} {1992})}\BibitemShut {NoStop}%
    \bibitem [{\citenamefont {Tiurev}\ \emph {et~al.}(2021)\citenamefont {Tiurev},
      \citenamefont {Mirambell}, \citenamefont {Lauritzen}, \citenamefont {Appel},
      \citenamefont {Tiranov}, \citenamefont {Lodahl},\ and\ \citenamefont
      {S\o{}rensen}}]{Tiurev2021}%
      \BibitemOpen
      \bibfield  {author} {\bibinfo {author} {\bibfnamefont {K.}~\bibnamefont
      {Tiurev}}, \bibinfo {author} {\bibfnamefont {P.~L.}\ \bibnamefont
      {Mirambell}}, \bibinfo {author} {\bibfnamefont {M.~B.}\ \bibnamefont
      {Lauritzen}}, \bibinfo {author} {\bibfnamefont {M.~H.}\ \bibnamefont
      {Appel}}, \bibinfo {author} {\bibfnamefont {A.}~\bibnamefont {Tiranov}},
      \bibinfo {author} {\bibfnamefont {P.}~\bibnamefont {Lodahl}},\ and\ \bibinfo
      {author} {\bibfnamefont {A.~S.}\ \bibnamefont {S\o{}rensen}},\ }\bibfield
      {title} {\bibinfo {title} {Fidelity of time-bin-entangled multiphoton states
      from a quantum emitter},\ }\href
      {https://doi.org/10.1103/PhysRevA.104.052604} {\bibfield  {journal} {\bibinfo
       {journal} {Phys. Rev. A}\ }\textbf {\bibinfo {volume} {104}},\ \bibinfo
      {pages} {052604} (\bibinfo {year} {2021})}\BibitemShut {NoStop}%
    \bibitem [{\citenamefont {S\o{}rensen}\ and\ \citenamefont
      {M\o{}lmer}(2003)}]{Anders2003}%
      \BibitemOpen
      \bibfield  {author} {\bibinfo {author} {\bibfnamefont {A.~S.}\ \bibnamefont
      {S\o{}rensen}}\ and\ \bibinfo {author} {\bibfnamefont {K.}~\bibnamefont
      {M\o{}lmer}},\ }\bibfield  {title} {\bibinfo {title} {Probabilistic
      generation of entanglement in optical cavities},\ }\href
      {https://doi.org/10.1103/PhysRevLett.90.127903} {\bibfield  {journal}
      {\bibinfo  {journal} {Phys. Rev. Lett.}\ }\textbf {\bibinfo {volume} {90}},\
      \bibinfo {pages} {127903} (\bibinfo {year} {2003})}\BibitemShut {NoStop}%
    \bibitem [{\citenamefont {T{\"o}rm{\"a}}\ \emph {et~al.}(1996)\citenamefont
      {T{\"o}rm{\"a}}, \citenamefont {Jex},\ and\ \citenamefont
      {Stenholm}}]{Torma1996}%
      \BibitemOpen
      \bibfield  {author} {\bibinfo {author} {\bibfnamefont {P.}~\bibnamefont
      {T{\"o}rm{\"a}}}, \bibinfo {author} {\bibfnamefont {I.}~\bibnamefont {Jex}},\
      and\ \bibinfo {author} {\bibfnamefont {S.}~\bibnamefont {Stenholm}},\
      }\bibfield  {title} {\bibinfo {title} {Beam splitter realizations of totally
      symmetric mode couplers},\ }\href {https://doi.org/10.1080/09500349608232738}
      {\bibfield  {journal} {\bibinfo  {journal} {Journal of Modern Optics}\
      }\textbf {\bibinfo {volume} {43}},\ \bibinfo {pages} {245} (\bibinfo {year}
      {1996})},\ \Eprint
      {https://arxiv.org/abs/https://doi.org/10.1080/09500349608232738}
      {https://doi.org/10.1080/09500349608232738} \BibitemShut {NoStop}%
    \bibitem [{\citenamefont {Carolan}\ \emph {et~al.}(2015)\citenamefont
      {Carolan}, \citenamefont {Harrold}, \citenamefont {Sparrow}, \citenamefont
      {Mart{\'\i}n-L{\'o}pez}, \citenamefont {Russell}, \citenamefont
      {Silverstone}, \citenamefont {Shadbolt}, \citenamefont {Matsuda},
      \citenamefont {Oguma}, \citenamefont {Itoh}, \citenamefont {Marshall},
      \citenamefont {Thompson}, \citenamefont {Matthews}, \citenamefont
      {Hashimoto}, \citenamefont {O'Brien},\ and\ \citenamefont
      {Laing}}]{Carolan2015}%
      \BibitemOpen
      \bibfield  {author} {\bibinfo {author} {\bibfnamefont {J.}~\bibnamefont
      {Carolan}}, \bibinfo {author} {\bibfnamefont {C.}~\bibnamefont {Harrold}},
      \bibinfo {author} {\bibfnamefont {C.}~\bibnamefont {Sparrow}}, \bibinfo
      {author} {\bibfnamefont {E.}~\bibnamefont {Mart{\'\i}n-L{\'o}pez}}, \bibinfo
      {author} {\bibfnamefont {N.~J.}\ \bibnamefont {Russell}}, \bibinfo {author}
      {\bibfnamefont {J.~W.}\ \bibnamefont {Silverstone}}, \bibinfo {author}
      {\bibfnamefont {P.~J.}\ \bibnamefont {Shadbolt}}, \bibinfo {author}
      {\bibfnamefont {N.}~\bibnamefont {Matsuda}}, \bibinfo {author} {\bibfnamefont
      {M.}~\bibnamefont {Oguma}}, \bibinfo {author} {\bibfnamefont
      {M.}~\bibnamefont {Itoh}}, \bibinfo {author} {\bibfnamefont {G.~D.}\
      \bibnamefont {Marshall}}, \bibinfo {author} {\bibfnamefont {M.~G.}\
      \bibnamefont {Thompson}}, \bibinfo {author} {\bibfnamefont {J.~C.~F.}\
      \bibnamefont {Matthews}}, \bibinfo {author} {\bibfnamefont {T.}~\bibnamefont
      {Hashimoto}}, \bibinfo {author} {\bibfnamefont {J.~L.}\ \bibnamefont
      {O'Brien}},\ and\ \bibinfo {author} {\bibfnamefont {A.}~\bibnamefont
      {Laing}},\ }\bibfield  {title} {\bibinfo {title} {Universal linear optics},\
      }\href {https://doi.org/10.1126/science.aab3642} {\bibfield  {journal}
      {\bibinfo  {journal} {Science}\ }\textbf {\bibinfo {volume} {349}},\ \bibinfo
      {pages} {711} (\bibinfo {year} {2015})},\ \Eprint
      {https://arxiv.org/abs/https://www.science.org/doi/pdf/10.1126/science.aab3642}
      {https://www.science.org/doi/pdf/10.1126/science.aab3642} \BibitemShut
      {NoStop}%
    \end{thebibliography}
    \end{document}